\documentclass{class_file}

\usepackage{amsmath}
\usepackage{mathtools}
\usepackage[T1]{fontenc}
\usepackage[latin1]{inputenc}
\usepackage{graphicx}
\usepackage{amssymb}
\usepackage{dcolumn}
\usepackage{color}
\usepackage{bm}
\usepackage[normalem]{ulem}
\usepackage{hyperref}
\usepackage[capitalise]{cleveref}
\usepackage[nice]{nicefrac}
\usepackage{url}
\usepackage[labelfont=bf]{caption}
\usepackage{verbatim}
\usepackage{soul}
\usepackage{pdfpages}
\usepackage{float}


\title{Half-integer thermal conductance in the absence of Majorana mode}

\author{Ujjal Roy$^{1,}$\footnote{These authors contributed equally: Ujjal Roy, Sourav Manna.}, Sourav Manna$^{2,*}$, Souvik Chakraborty$^{1}$, Kenji Watanabe$^{3}$, Takashi Taniguchi$^{3}$, Ankur Das$^{4}$, Moshe Goldstein$^{5}$, Yuval Gefen$^{2}$ and Anindya Das$^{1}$\footnote{anindya@iisc.ac.in}}

\begin{document}

\maketitle

\begin{affiliations}
\item Department of Physics, Indian Institute of Science, Bangalore, 560012, India.
\item Department of Condensed Matter Physics, Weizmann Institute of Science, Rehovot 76100, Israel.
\item National Institute of Material Science, 1-1 Namiki, Tsukuba 305-0044, Japan.
\item Department of Physics, Indian Institute of Science Education and Research (IISER) Tirupati, Tirupati 517619, India.
\item Raymond and Beverly Sackler School of Physics and Astronomy, Tel-Aviv University, Tel Aviv, 6997801, Israel.
\end{affiliations}
\begin{abstract}

\noindent\textbf{Considering a range of candidate quantum  phases of matter, half-integer thermal
conductance ($\kappa_{\text{th}}$) is believed to be an unambiguous evidence of non-Abelian states.
It has been long known that such half-integer values arise due to the presence of Majorana
edge modes, representing a significant step towards topological quantum computing
platforms. Here, we challenge this prevailing notion by presenting a comprehensive
theoretical and experimental study where half-integer two-terminal thermal conductance
plateau is realized employing Abelian phases. Our proposed setup features a confined
geometry of bilayer graphene, interfacing distinct particle-like and hole-like integer quantum
Hall states. Each segment of the device exhibits full charge and thermal equilibration. Our approach is amenable to generalization to other quantum Hall platforms, and may give rise to other values of fractional (electrical and thermal) quantized transport.
Our study demonstrates that the observation of robust non-integer values of thermal conductance can arise as
a manifestation of mundane equilibration dynamics
as opposed to underlying non-trivial topology.}

\end{abstract}

\noindent\textbf{Introduction.}
Non-Abelian phases of matter---ranging from fractional quantum Hall (FQH) states \cite{Moore-Read1} to quantum spin liquids\cite{Matsuda2025-qc}---are at the forefront of strongly-correlated quantum many-body
physics. Such phases are potential host of exotic
quasiparticles exhibiting non-Abelian  braiding statistics, which are proposed ingredients for
topological quantum computation \cite{arxiv.2210.10255}.
Half-integer or even-denominator FQH states \cite{Ma2024-qr} are a fertile playground
for hosting such quasiparticles, thereby being a focal point of research  over the past decades. The most appealing non-Abelian bulk phases \cite{Ma2024-qr}
are the Pfaffian, its hole-conjugate partner---the anti-Pfaffian, and the
particle-hole symmetric Pfaffian, discarding
previously proposed Abelian platforms. Among the various experimental probes, thermal conductance offers profound insights into these phases \cite{Banerjee2018,Paul2024}. Whether dealing with bosons, simple integer quantum Hall states, or Abelian anyons \cite{Kane1996-us,Kane1997-dd,Greiner1997-pk,Pierre2013,Banerjee2017,Srivastav2019,Srivastav2021,Srivastav2022}--- classified into simple particle-like and complex hole-like (with upstream neutral modes) FQH liquids --- thermal conductance is consistently quantized to an integer value in the units of $\kappa_0T$, $\kappa_0=\frac{\pi^2k_B^2}{3h}$, where $k_B$ is the Boltzmann constant, $h$ is the Planck's constant, and $T$ being the temperature. By contrast, a half-integer quantization of thermal conductance, $\frac{1}{2}\kappa_0 T$, has been taken as a definite indicator of non-Abelian phases \cite{Ma2024-qr,Matsuda2025-qc}, serving as compelling evidence for the presence of Majorana edge mode\cite{Banerjee2018,Paul2024}. Importantly, as has been demonstrated by earlier works\cite{Ma2024-qr}, half integer values of the electrical conductance can emerge due to dynamics extraneous to the Abelian or non-Abelian nature of the underlying FQH phase.\\\\
A natural follow-up question is whether half-integer conductance values are exclusive to even-denominator FQH states. Interestingly, theoretical work \cite{Abanin2007-me} has proposed that fractional electrical conductance values can be engineered in graphene n-p or n-p-n junctions using integer QH states. Experimentally, both fractional and half-integer electrical conductance values have been observed in such systems \cite{Williams2007-zo,Kim_npn,Zimmermann2017-dd,Pandey2024}. Moreover, half-integer electrical conductance has been realized in other platforms as well, such as quantum anomalous Hall insulator-superconductor junctions \cite{wen2018prl,Kayyalha2020-we,uday2025}, and in quantum point contact (QPC) responses at $\nu = 2/3$ in GaAs-based FQH systems \cite{manfra_2/3,Fauzi2023-ki}. It is important to emphasize that these observations are not associated with non-Abelian bulk states but rather represent engineered half-integer electrical conductance in systems governed by Abelian physics. This raises a fundamental question: for such platforms that mimic fractional conductance plateaus,
can one also expect a half-integer quantization of thermal conductance, $\frac{1}{2}\kappa_0 T$, which is widely believed to be a hallmark of non-Abelian states. As of now, no theoretical or experimental studies have definitively addressed this question.\\\\
\begin{figure}[htbp]
\centerline{\includegraphics[width=0.85\textwidth]{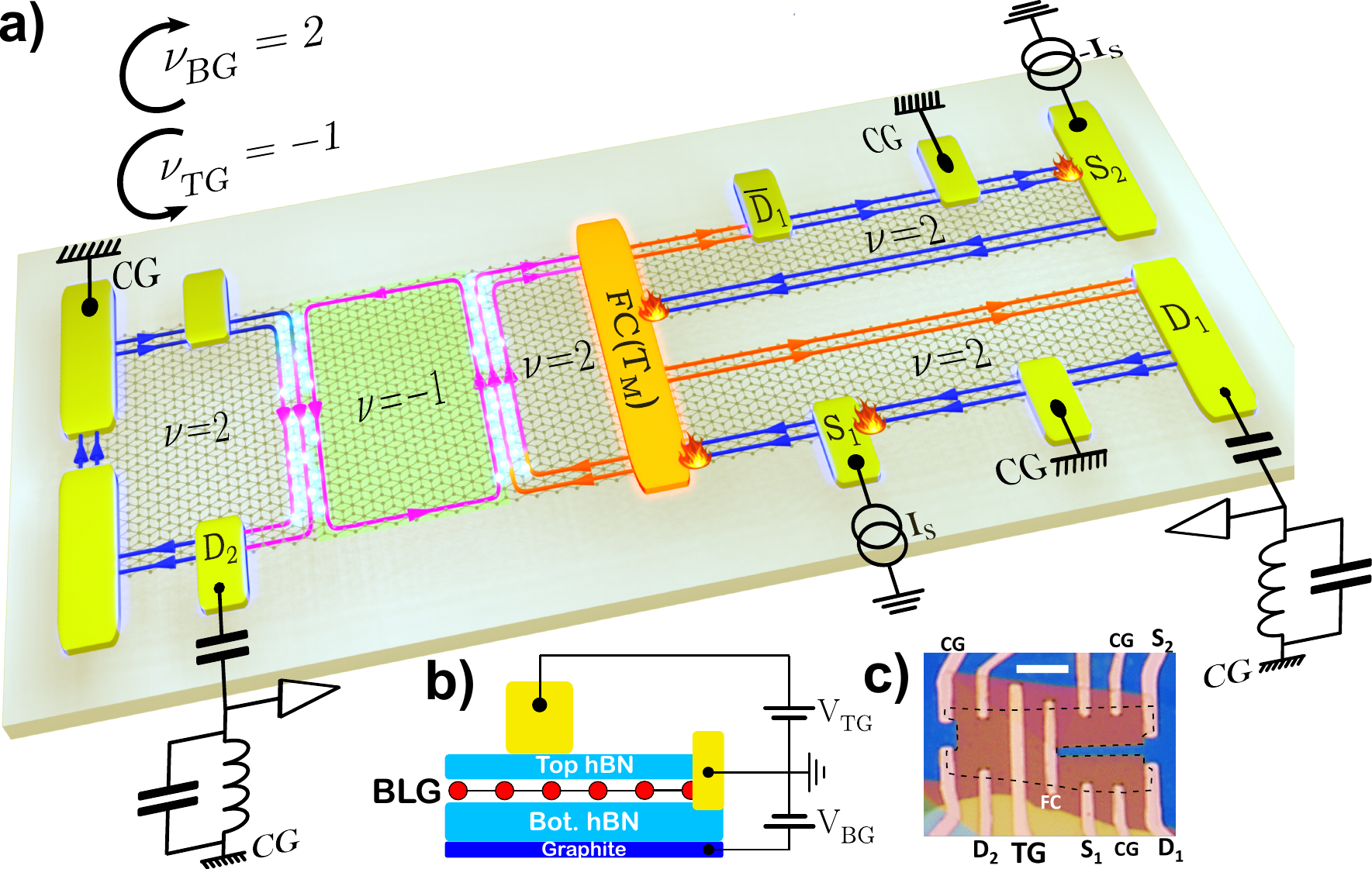}}
\caption{\textbf{Device schematic, cross-section and optical image.} \textbf{(a)} Device schematic and measurement set-up (see SI section 1 for more details). The device consists of three BLG arms connected by a central floating contact (FC), of which two identical arms on the right hand side are kept at $\nu=2$, controlled by global graphite back-gate and the other one having a local top-gate (green shaded region) in the middle with $\nu=-1$ forming an n-p-n segment with two interfaces of co-propagating electron-hole chiral-edges. The direction of the applied out of plane magnetic field dictates the chirality of the electron and hole edges, shown by the arrows associated with each chirals. The full charge and heat-equilibration along the length of the two interfaces is illustrated by the white glowing spheres. Each of the arms has atleast one cold ground contact labelled as CG. To measure the thermal conductance, noiseless dc currents $I_{S}$ and $-I_{S}$ are injected simultaneously at $S_1$ and $S_2$ respectively, which increases the electronic temperature of FC to $T_M$, while keeping it's chemical potential (or voltage) to zero, and to determine the $T_M$, Johnson-Nyquist noise is measured both at $D_1$ and $D_2$, with the help of resonant LC network. The color of each chiral signifies it's temperature. For the chirals emerging from FC are the hottest, shown by orange. Then after equilibration it reduces it's temperature and hence shown by pink. The coldest ones are shown by blue, thermalized to temperature $T_0$. The hot-spot positions are marked by the fire signs which indicate the changing points for the chemical potential (or DC voltage) along the chirals. \textbf{(b)} Device cross-section showing the BLG is encapsulated by the top and bottom hBN. The global graphite back-gate and metallic local top-gate is connected external voltage source $V_{BG}$ and $V_{TG}$ respectively, which controls the respective filling factors. \textbf{(c)} Optical image of the device with the individual contacts marked same as in \textbf{(a)}. The size of the scale bar is $5\mu m$.}
\label{Device sch.}
\end{figure}
In this work, we demonstrate that a robust half-integer thermal conductance can be realized in a complex structure, which is not subject to fine-tuning of the system's parameters. Such half-integer values are not a manifestation of an underlying non-Abelian topology, and in particular they do not represent an underlying Majorana mode. We utilize n-p-n heterojunctions consist of symmetry broken $\nu'=-1$ region sandwiched between two $\nu=2$ integer QH state in the zeroth Landau level (ZLL) of bilayer graphene (BLG) to artificially create a $\frac{1}{2}G_0$ electrical conductance state, where $G_0=e^2/h$, is the electrical conductance quanta. The high tunability of symmetry-broken flavors (spin, valley, and orbital)\cite{McCann2006,McCann_2013,Maher2014-gg,Zhu2018} in the ZLL of BLG --- through both magnetic and electric fields --- enables full equilibration between the co-propagating electron-hole (e-h) modes for which the two-terminal electrical conductance becomes, $G_{\nu-\nu'-\nu}=\frac{|\nu||\nu'|}{|\nu|+2|\nu'|}G_0=\frac{1}{2}G_0$\cite{Kim_npn} (see "Methods" and supplementary). To test the idea that, two-terminal thermal conductance of such QH n-p-n junction also leads to a value of $\frac{1}{2}\kappa_0T$ (which is proposed by our theory as discussed in method and Supplementary Information (SI) section 12), we employ a unique device geometry with three arms as shown schematically in \cref{Device sch.}a. This geometry enables the creation of a temperature difference ($\Delta T$) across the n-p-n junction while maintaining zero net charge current, despite heat transport via chaotic (e-h) mixing across the junction. Our data suggests, full equilibration of both charge and energy leading to thermal conductance of $\sim \frac{1}{2}\kappa_0T$, consistent with our theoretical calculations. This challenges our current understanding as half-quantization of heat conductance is a unique signature of a Majorana mode, which makes this work is of fundamental importance. Our results also confirm that the Wiedemann-Franz law remains valid for engineered fractional values of thermal conductance, in contrast to the behavior observed in anyonic heat flow \cite{Banerjee2017,Banerjee2018,Srivastav2021,Srivastav2022}. We reproduce this result in another device consists of a p-n-p junction with $\nu=-2$ and $\nu'=1$, reinforcing the reliability of our findings.

\noindent\textbf{Device and working principle.} We use high-quality hBN encapsulated BLG heterostructures with graphite as the global back gate (BG) (see "Methods" for device fabrication details).
As shown in \cref{Device sch.}a, the schematic of the device consists of three arms with a central floating contact (FC).
To the right of the FC, there are two identical BLG channels with densities controlled by global BG. In contrast, in the left channel, a local top gate (TG) is placed over a specific region to create a n-p-n or p-n-p heterojunction.
In this configuration, co-propagating electron and hole QH edge modes move along the interface of the n-p (and p-n) region, and their respective filling factors, $\nu$ and $\nu'$ are tuned by $V_{BG}$ and $V_{TG}$. Here, our chosen platform BLG plays a crucial role in this experiment --- not only it allows tuning between electron or hole carriers by adjusting the gate voltages, but it also facilitates equilibration between the e-h modes due to the tunability of symmetry-broken flavors (spin, valley, and orbital) of ZLL of BLG with magnetic and displacement field\cite{Zhu2018}. In the SI section 5, we show
how the flavors of symmetry-broken edge modes of ZLL of BLG at the n-p (and p-n) interface depend on the displacement fields of the global (controlled by only BG) and local part (controlled by both BG and TG).
If the flavors are same at the interface, one would expect full equilibration of chemical potential and energy, and the two-terminal electrical conductance across the junction is expected to be $G_{\nu-\nu'-\nu}=\frac{|\nu||\nu'|}{|\nu|+2|\nu'|}G_0$\cite{Kim_npn} (see "Methods" for the theoretical calculation). For $\nu=2$ and $\nu'=-1$, the $G_{\nu-\nu'-\nu}$ expected to be $\frac{1}{2}G_0$, thus one can engineer half-integer electrical conductance. To measure thermal conductance across the n-p-n junction, our three-arm device (\cref{Device sch.}a) has the following advantages in contrast to our earlier works on two-arm devices\cite{Srivastav2019,Srivastav2021,Srivastav2022,Kumar2024}. In order to create the temperature difference across the n-p-n junction, simultaneously, an electron and a hole current of equal magnitude are injected to the FC (from the identical two arms placed to the right side of the FC in \cref{Device sch.}a) to maintain its effective chemical potential at zero. This ensures no voltage drop across the n-p-n junction while having hotspots at the FC to increase its temperature as shown schematically in \cref{Device sch.}a. This is crucial because electrically biased graphene n-p-n juction would otherwise generate shot noise due to the current partitioning of co-propagating e-h modes at the n-p (and p-n) interface\cite{Matsuo2015-lk,Paul2020-ax}, overshadowing the Johnson-Nyquist signal carrying the $\Delta T$ information. This device geometry enables us to overcome this experimental hurdle by eliminating shot noise as there is no voltage difference across the junction and therefore allows us to successfully measure the $\Delta T$ even though heat is carried through the chaotic e-h mixing across the n-p-n junction. For full thermal equilibration, the thermal conductance of such n-p-n junction will be, $K_{\nu-\nu'-\nu}=\frac{|\nu||\nu'|}{|\nu|+2|\nu'|}k_0 T$ (see "Methods" for the theoretical calculation) and for $\nu=2$ and $\nu'=-1$, the $K_{\nu-\nu'-\nu}$ expected to be $\frac{1}{2}\kappa_0T$ --- a half-integer thermal conductance.

\begin{figure*}[htbp]
\centerline{\includegraphics[width=0.9\textwidth]{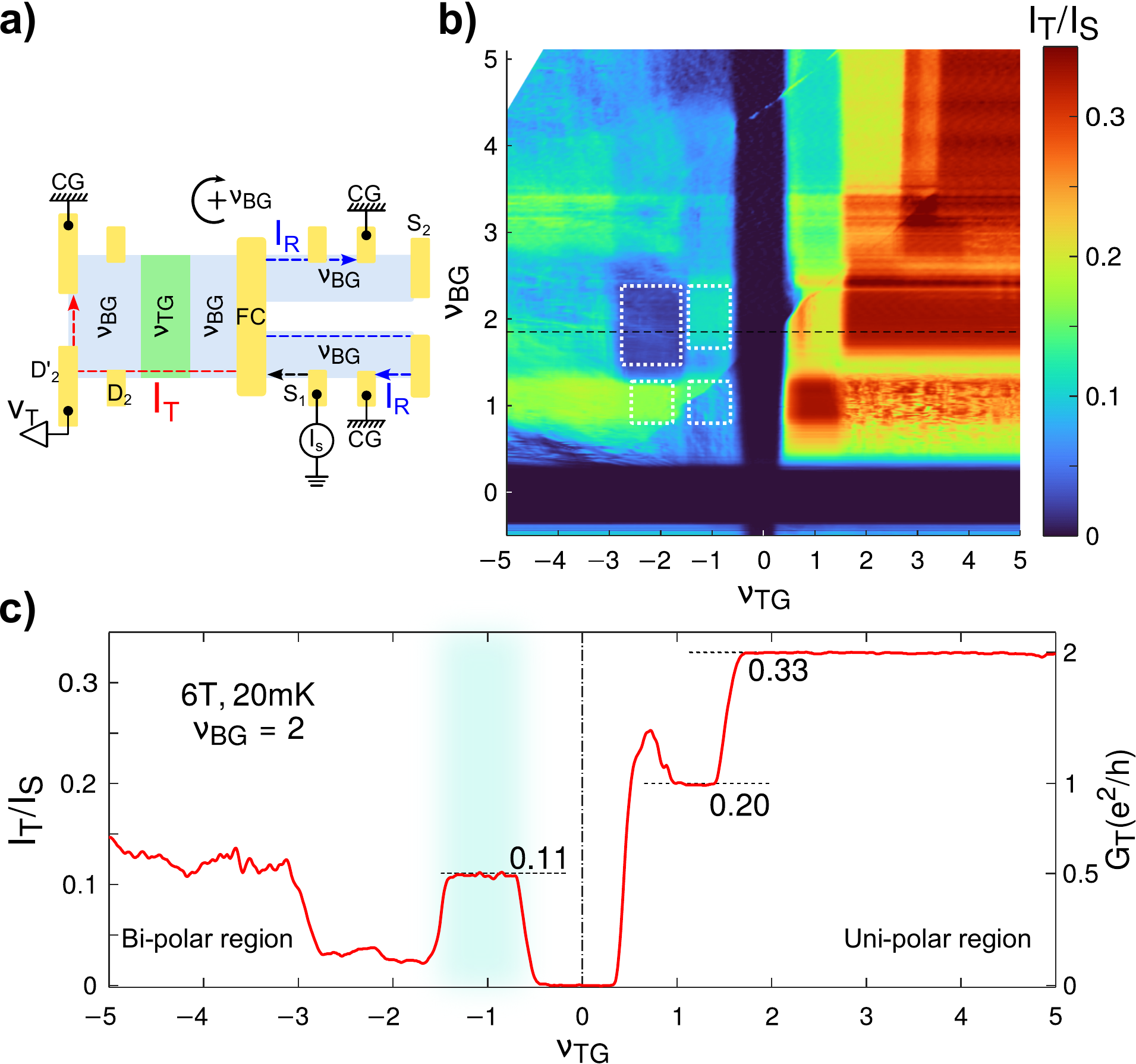}}
\caption{\textbf{Device charaterzation.} \textbf{(a)} Sketch of the measurement setup used to probe device response, using standard lock-in techniques. A current $I_{S}\sim5nA$ is injected to either $S_1$ or $S_2$ while ensuring that at least one ground is present in each arm. The magnetic field direction is chosen such that, chirality of electronic carriers remain clock-wise. Black dashed line represents $I_S$ reaching the FC, where it splits into three parts. Two equal portions, labeled $I_R$ (blue dashed line) propagate along the right side of FC and terminate into respective CG, while the remaining one $I_T$ (red dashed line), travels along the left of FC, which is given by $I_T=V_T|\nu_{BG}|G_0$, where $V_T$ is the measured voltage drop at $D_2'$ (see SI section 4 for more details). \textbf{(b)} 2D colormap of the measured $I_T$ normalized to $I_S$ as a function of $\nu_{BG}$ and $\nu_{TG}$. \textbf{(c)} Line cut along the black dashed-line in \textbf{(b)}, showing the variation of $I_T/I_S$ with $\nu_{TG}$ while the bulk filling factor remains at $2$. Highlighted region around $\nu_{TG}=-1$ showing $I_T/I_S=0.11$ corresponding to $G_T=0.5G_0$. The vertical line divides at $\nu_{TG}=0$, separates the uni-polar and bi-polar region. The values of $I_T/I_S$ at other marked plateau regions, e.g., $\nu_{TG}=2$ and $1$, follows $G_{T}=min(|\nu_{BG}|,|\nu_{TG}|)G_0$.}
\label{Device Char.}
\end{figure*}

\noindent\textbf{Results:}
\\
\noindent\textbf{Electrical conductance.} To measure the electrical conductance across the n-p-n junction of our device geometry in \cref{Device sch.}a, we use the standard lock-in-based measurements with ac excitation frequency of $\sim 13 Hz$. All the measurements are done in a cryo-free dilution fridge at base temperature, $T \sim 20mK$ (see "Methods"). After confirming the presence of well-developed symmetry-broken QH states at each arm (see SI section 3) of our device, we inject a constant current $I_{S} \sim5nA$ at $S_1$ (or $S_2$) and measure the transmitted current through the n-p-n region (the top-gated part of the left arm of the device) in \cref{Device sch.}a. Upon reaching the FC, the injected current, $I_S$ divides into three parts, following to the conductance of each arm. These currents flow along the device boundaries and terminate at the respective grounds as shown schematically in \cref{Device Char.}a, which can also be understood in the simplified conductor model as shown in SI section 4. The conductance of the two identical arms on the right of the FC are same, and is given by $G_R=|\nu|G_0$, with corresponding current $I_R$. Whereas, for the left arm, the current, $I_T$ depends on the trans-conductance of the n-p-n junction, which follows $G_{T}=min(|\nu|,|\nu'|)G_0$ in the uni-polar regime, and for the bi-polar regime, it is given by $G_{T}=\frac{|\nu||\nu'|}{|\nu|+2|\nu'|}G_0$ for the case of full charge equilibration of the co-propagating edges at the n-p (and p-n) interface. The current conservation follows as $I_S=2I_R+I_T$ and the ratio of $\frac{I_T}{I_S}=\frac{G_T}{2G_R + G_T}$. Since, $G_R$ is known, hence, by measuring the transmitted current $I_T$ through the n-p-n junction, one can infer the $G_T = G_{\nu-\nu'-\nu}$. The details about $I_T$ measurement are discussed in the SI section 4.

The measured $I_T/I_S$ as a function of $\nu_{BG}$ ($\nu$) and $\nu_{TG}$ ($\nu'$) is shown as a 2D colormap in \cref{Device Char.}b for 6T magnetic field. The vertical dark blue strip around $\nu_{TG}=0$, where $I_T=0$, indicates the top-gated region becomes insulating, making the boundary between the uni-polar and bi-polar region. Around $\nu_{BG} = 0$ we also observe $I_T=0$ as the entire bulk becomes insulating. The right-half of the plot, representing the uni-polar region, is divided into rectangular blocks of different colors centralized around the integer values of $\nu_{BG}$ and $\nu_{TG}$ signifying the robust plateaus of $I_T$ at different filling configurations as expected. For instance, for $\nu_{BG}=\nu_{TG}$, the injected current $I_S$ is expected to divide equally among the three arms leading to $\frac{I_T}{I_S}=0.33$. This is indeed observed in the measured data for $(\nu_{BG},\nu_{TG})=(1,1)$, $(2,2)$, and $(4,4)$, and also close to the expected value for $(3,3)$ as well. This also satisfies an essential requirement for thermal conductance measurements -- equipartition of current at FC, and will be discussed later in details (see the simplified conductor model illustrated in SI section 4). However, in the bi-polar region (left-half portion), the $\frac{I_T}{I_S}$ is not simple for different combinations of ($\nu_{BG}, -\nu_{TG}$) since its value depends on the degree of equilibration of the co-propagating edges at the n-p (and p-n) interface. It can be seen from \cref{Device Char.}b (white dotted boxes) that for (2,-1) and (1,-2)
the measured $I_T$ suggests robust plateau with full equilibration ($G_{\nu-\nu'-\nu}=\frac{|\nu||\nu'|}{|\nu|+2|\nu'|}G_0$) in contrast to (1, -1) and (2,-2)
with partial equilibration (measured value $< \frac{|\nu||\nu'|}{|\nu|+2|\nu'|}G_0$). The full and partial equilibration can be understood in terms of the spin state of the co-propagating edges at the interface, as discussed in the SI section 5 based on prior studies\cite{Maher2013-ej,Zhu2018}, and also present the data with various magnetic fields revealing a clear transition of (2,-1) from partial to full equilibration.
In \cref{Device Char.}c, we show the trace of $\frac{I_T}{I_S}$ as a function of $\nu_{TG}$ while $\nu_{BG} = 2$. It can be seen for (2,-1), $\frac{I_T}{I_S} \sim 0.11$ with robust plateau indicating $G_{\nu-\nu'-\nu}=\frac{|\nu||\nu'|}{|\nu|+2|\nu'|}G_0 = 0.5G_0$. This is repeated in another device, as shown in SI section 9. Though, the equilibration for many combinations of filling factors can be understood in terms of the spin state of the co-propagating edges, however, anomalously suppressed equilibration observed for (2,-2) requires further theoretical understanding beyond the spin-selective equilibration model (see SI section 5 for the discussion).

\begin{figure*}[htbp]
\centerline{\includegraphics[width=0.95\textwidth]{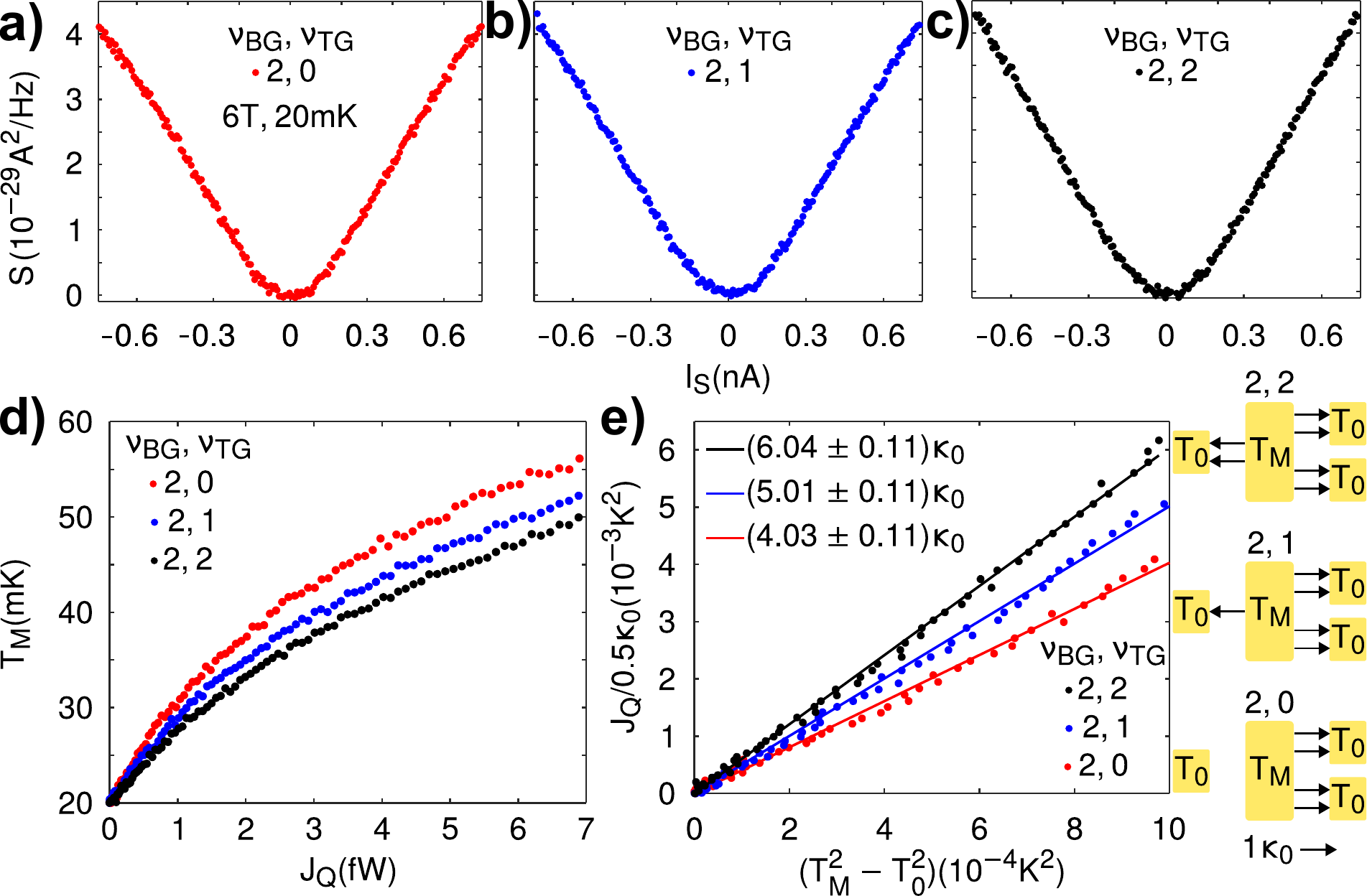}}
\caption{\textbf{Thermal-Hall conductance in the uni-polar region.} \textbf{(a)-(c)} Excess thermal noise is measured at $D_1$ as a function of injected dc current $I_{S}$ for different configuration of $(\nu_{BG},\nu_{TG})=(2,0)$ \textbf{(a)}, $(2,1)$ \textbf{(b)} and $(2,2)$ \textbf{(c)}. \textbf{(d)} The increased temperatures $T_M$ of floating contact, are extracted from panel \textbf{(a)-(c)}, are plotted (solid circles) against the dissipated power $J_Q$($=P$), due to Joule heating in FC. \textbf{(e)} $J_Q/0.5\kappa_0$ is plotted (solid circles) as a function of $(T_M^2-T_0^2)$ for $(2,2)$, $(2,1)$ and $(2,0)$. The solid lines are the linear fittings to extract the thermal conductance, parameterized as $N\kappa_0$ (\cref{heat_bal}), with $N$ as the of outgoing channels. Slope of this fits give $N$ as $6.04$, $5.01$ and $4.03$, respectively as expected. A simple diagram of the heat-flow depicted in the right. From this we can deduce $N=2|\nu_{BG}|+min(|\nu_{BG}|,|\nu_{TG}|)$.}
\label{heat_cond_uni}
\end{figure*}

\noindent\textbf{Thermal conductance.} To measure the thermal conductance, we simultaneously inject dc currents, $I_S$ and $-I_S$ into FC from sources $S_1$ and $S_2$, respectively, as shown in \cref{Device sch.}a to maintain the effective chemical potential of FC at zero. In this configuration, the power dissipation at FC is given by $P=\frac{I_S^2}{\nu_{BG}G_0}$\cite{Srivastav2019,Srivastav2021,Srivastav2022,Kumar2024} (See "Methods" for details). This results in an increase in the electronic temperature of the FC and, its steady-state temperature, $T_M$ is determined by the following heat balance relation\cite{Pierre2013}:
\begin{equation}
P=J_Q=J_Q^e(T_M,T_0)+J_Q^{e-ph}(T_M,T_0)=0.5N\kappa_0(T_M^2-T_0^2)+J_Q^{e-ph}
\label{heat_bal}
\end{equation}
Here, $J_Q^e(T_M,T_0)$ represents the electronic contribution of the heat current via $N$ outgoing chiral channels from the FC, $J_Q^{e-ph}$ accounts for the heat loss due to electron-phonon cooling, and $T_0$ is the electron temperature of the cold reservoirs. To obtain $T_M$, we measure the excess thermal noise at detectors $D_1$ (or $\bar{D_1}$) and $D_2$, positioned along the outgoing edge channel and using the Nyquist-Johnson relation\cite{Pierre2013,Banerjee2017}, $S=2G^*k_B(T_M-T_0)$ for uni-polar case (i.e., when $\nu_{BG}$ and $\nu_{TG}$ have the same sign). Here, $\frac{1}{G^*}=\frac{1}{G_{amp}}+\frac{1}{\sum_{i=1,i\neq amp}^{3}{G_i}}$, $G_{amp}$ is the conductance of the arm with the detector, and $G_i$ represents the conductance of other arms. For the right arms in \cref{Device sch.}a, the conductance, $G_{i}=\nu_{BG}G_0$,
while for the left arm, $G_{i}=min(\nu_{BG},\nu_{TG})G_0$.

Before we present the thermal conductance data for the bi-polar regime,
we first benchmark the thermal conductance values
for the uni-polar region.
As shown in \cref{Device Char.}c, we see robust plateaus around $\nu_{TG}=0$, $1$ and $2$, while the bulk filling is kept at $\nu_{BG}=2$.
We measure the excess noise, $S_I$, at the detector $D_1$ for (2,0), (2,1) and (2,2), and shown in \cref{heat_cond_uni}a-c, as a function of injected dc current $I_S$ (and $-I_S$). The extracted $T_M$ from the $S_I$ for different $(\nu_{BG},\nu_{TG})$ are shown in \cref{heat_cond_uni}d, against the dissipated power, $J_Q$. The electronic temperature $T_0$ is $\approx20mK$ (its determination is shown in SI section 2).
In \cref{heat_cond_uni}e,
the $J_Q/0.5\kappa_0$ is plotted with $(T_M^2-T_0^2)$. The linear fitting of each plot will give the value of the thermal conductance, and the extracted values are $6.04 \kappa_0$, $5.01 \kappa_0$ and $4.03 \kappa_0$, respectively, for $(\nu_{BG},\nu_{TG}) = (2,2)$, $(2,1)$ and $(2,0)$. The linearity of each plot suggests a negligible e-ph contribution (second term in \cref{heat_bal}) within the displayed range of $T_M\approx50mK$.
Our results are consistent with the simplified heat-flow diagram shown in the right of \cref{heat_cond_uni}e: when $\nu_{BG}=2$, two identical right arms of the FC always hosts total $4$ number of channels, while the number of channels, on the left arm of the FC, are varied from $2$ to zero by tuning $\nu_{TG}$, thus, in general, one can express the total number of out-going channels ($N$) from the FC as $N=2|\nu_{BG}|+min(|\nu_{BG}|,|\nu_{TG}|)$.

\begin{figure*}[htbp]
\centerline{\includegraphics[width=0.9\textwidth]{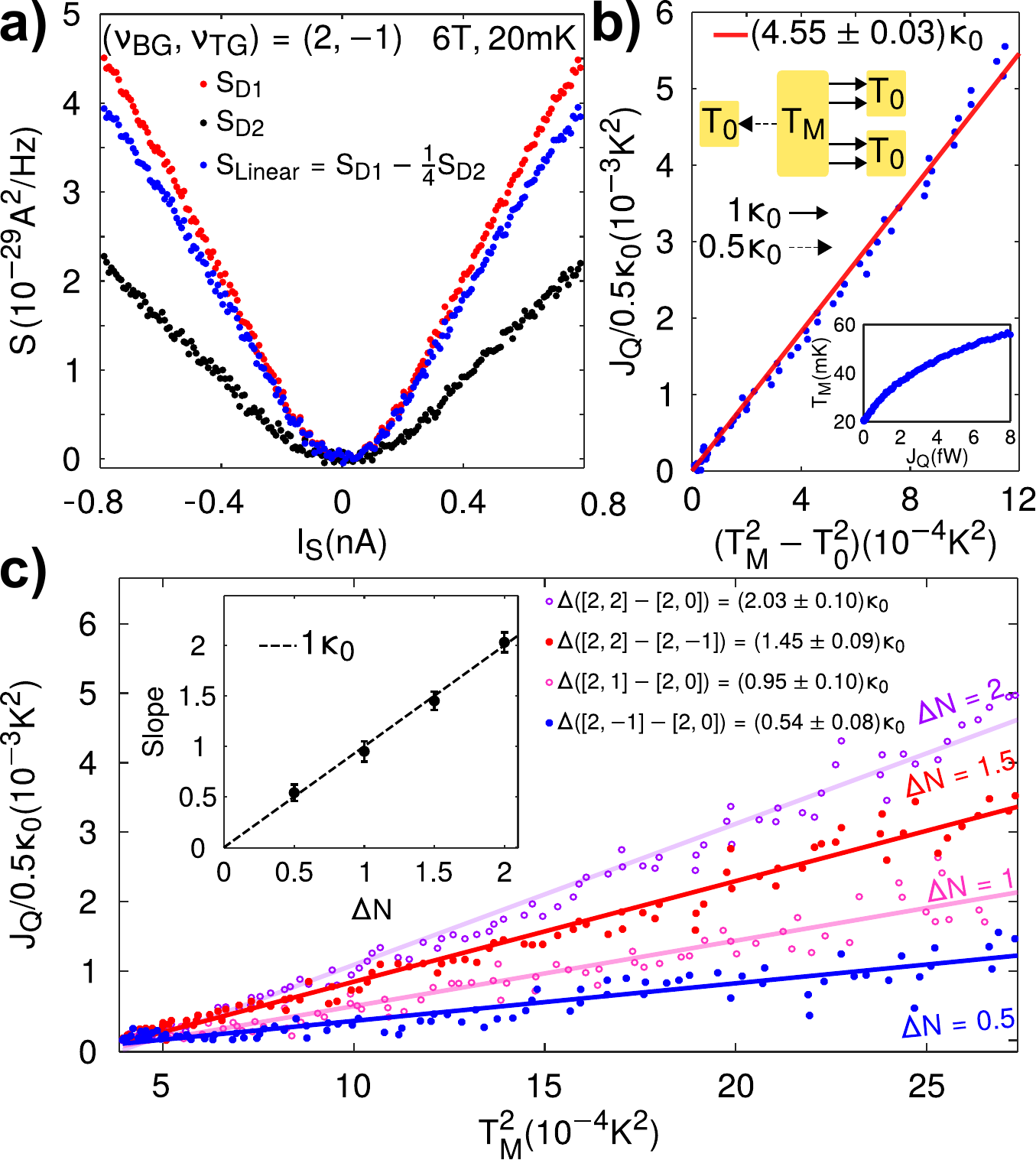}}
\caption{\textbf{Half-integer thermal conductance.} \textbf{(a)} Excess thermal noise measured at the half-conductance plateau $(\nu_{BG},\nu_{TG})=(2,-1)$, at $D_1$ (red circles, $S_{D1}$) and at $D_2$ (black Circles, $S_{D2}$) as a function of $I_S$. $S_{Linear}=S_{D1}-\frac{1}{4}S_{D2}$ is shown by the blue circles. \textbf{(b)} Inset: Increased temperature of the floating contact $T_M$ extracted from $S_{Linear}$ in panel \textbf{(a)} is plotted against the dissipated power $J_Q$. $\Delta J_Q/0.5\kappa_0$ is plotted (solid circles) as a function of $T_M^2-T_0^2$. The slope of the linear fitting (solid line) gives the thermal conductance value which is expected to be $4.5\kappa_0$, which is depicted by the heat-flow diagram having $4$-channel of heat conductance of $\kappa_0$ and one effective channel with net thermal conductance of $0.5\kappa_0$. \textbf{(c)} $\Delta J_Q/0.5\kappa_0$ is plotted against $T_M^2$ for $\Delta N=2$ (between $[2,2]$ \& $[2,0]$), $\Delta N=1.5$ (between $[2,2]$ \& $[2,-1]$), $\Delta N=1$ (between $[2,2]$ \& $[2,1]$) and $\Delta N=0.5$ (between $[2,-1]$ \& $[2,0]$), where $\Delta J_Q=J_Q(N_i,T_M)-J_Q(N_j,T_M)$. The solid lines are linear fittings to extract the the thermal conductance associated with $\Delta N$ $1D$ channels. Inset shows the slopes extracted from each solid lines vs $\Delta N$. The error bar represents the uncertainty of the fitting. The dashed line shows the theoretical $1\kappa_0$ line.}
\label{half_heat_cond}
\end{figure*}

\noindent\textbf{Half-integer thermal conductance.} In this section, we present the thermal conductance for the bi-polar regime of $(\nu_{BG},\nu_{TG}) = (2,-1)$.
However, generically in the bi-polar regime, the noise \cite{PhysRevLett.132.136502,SMlong,SMshort,SMAD} to temperature conversion is not simple and requires solving higher-order equations (see the "Methods" and the SI section 12 for details).
A simplified approach is to measure the excess noise simultaneously both at $D_1$(or $\bar{D_1}$) and $D_2$, leading to a linear relation (see SI section 12 for details):
\begin{equation}
S_{Linear}=S_{D1}-\frac{1}{4}S_{D2}=|\nu_{BG}|G_0k_B(T_M-T_0).
\label{si_bipolar}
\end{equation}
For the (2,-1) plateau, the excess thermal noise, $S_{D1}$ (red circles) and $S_{D2}$ (black circles) measured at $D_1$ and $D_2$, respectively, are shown in Fig. 4a as a function of $I_S$ (and $-I_S$). The $S_{Linear}$ is shown by the blue circles in \cref{half_heat_cond}a, which is further converted to $T_M$ using \cref{si_bipolar} and shown as a function of $J_Q$ in the lower inset of \cref{half_heat_cond}b. The $J_Q/0.5\kappa_0$ is plotted (blue circles) as a function of $T_M^2-T_0^2$ in \cref{half_heat_cond}b and the linear fitting gives the value of the thermal conductance $\sim 4.55\kappa_0$, which is depicted by the heat-flow diagram having $4$ out-going channels via the two identical right arms (in \cref{Device sch.}a), and an effective channel with net thermal conductance of $\sim 0.5\kappa_0$ through the left arm consists of the n-p-n junction ($(\nu_{BG},\nu_{TG}) = (2,-1)$). This suggests the thermal conductance of the n-p-n junction is $\sim 0.5\kappa_0$. To further confirm it, we plot the $\Delta J_Q/0.5\kappa_0$ with $T_M^2$ for different configurations of $\Delta\nu$: $\Delta$([2,2] - [2,0]), $\Delta$([2,2] - [2,-1]), $\Delta$([2,2] - [2,1]) and $\Delta$([2,-1] - [2,0]) in \cref{half_heat_cond}c, and the corresponding linear fittings give the thermal conductance values of $\sim 2.03\kappa_0$, $1.45\kappa_0$, $0.95\kappa_0$, $0.54\kappa_0$, respectively. As expected for $\Delta$([2,2] - [2,0]) and $\Delta$([2,2] - [2,1]), the thermal conductance values are consistent with the effective number of channels $\Delta N=2$ and $\Delta N=1$, respectively. However, $\Delta$([2,2] - [2,-1]) with $\sim 1.45\kappa_0$ and $\Delta$([2,-1] - [2,0]) with $\sim 0.54\kappa_0$ reconfirms the half-integer thermal conductance of the n-p-n junction with the effective number of channels $\Delta N=0.5$. This is further reemphasized in \cref{half_heat_cond}c inset via plotting the extracted thermal conductance values with $\Delta N$, and we see a linear increment with the slope of $1\kappa_0$. Similar observations are presented for Device 2 in Supplementary Information Section 11.

\noindent\textbf{Discussion and outlook.}
The observation reported here of a plateau of half-integer thermal (as well as electrical) conductance relies not on underlying topology characterizing non-Abelian phases, but rather on the robustness of fully-equilibrated edge modes to local perturbations. Specifically, we address a "bi-polar" boundary, (2,-1), separating two integer quantum Hall phases. The feasibility of inter-mode full equilibration in our study depends on the flavors of symmetry-broken edge modes as well as on the presence of disorder\cite{DGG2014,Long2008,Li2008,Abanin2007-me} at the bipolar junction. Our device, having a boundary of length $\sim 10\mu m$, an order of magnitude larger than earlier works\cite{DGG2014,Nam2011-di,Morikawa2015-vs}, allows full equilibration of edge modes. The resulting edge transport is expected to be incoherent for our extended transport path, consistent with earlier findings\cite{Paul2020-ax}, where the measured shot noise Fano distinguishes the coherent versus incoherent process at the bipolar regime.

Our work puts on the table the issue of robustness of quantum transport platforms to external perturbations. One may compare topology-based resilience to resilience underlined by strong equilibration dynamics. Such a study can use case studies of other equilibration-engineered fractional values of thermal and electrical conductance, as well as also serving a test-bed for the validity of the Wiedemann-Franz law. This might be particularly interesting when implemented to platforms which (in the absence of equilibration) exhibit charge fractionalizaton giving rise, e.g., to charged and neutral eigenmodes \cite{PhysRevLett.102.236402, Bid2010,kumar_2022,Kumar_2024magnon}. Generalizations to fractional quantum Hall boundaries (e.g., with charge and neutral modes \cite{PhysRevLett.111.246803,kumar_2022} or charge and spin modes \cite{PhysRevB.55.7818}) should follow. This work also opens up an intriguing direction of incorporating equilibration dynamics (e.g. \cite{PhysRevLett.134.096303}) into the physics of dilute colliding beams\cite{Zhang2024,collider,Han2016,PhysRevLett.116.156802}, involving both fermions and anyons. We also note that half-integer thermal conductance that does not involve non-Abelian modes may appear in other contexts\cite{Nicholas_2023}.

\textit{Note added--} While completing this draft, we came to know of a recent manuscript by Karmakar et al.: arXiv: 2505.08746 (2025).

\noindent\textbf{Methods}
\\
\noindent\textbf{Device fabrication.} Our devices consists of Van-der Waals heterostructure of hBN/BLG/hBN/Graphite on a p-doped Si substrate with $285nm$ of $SiO_2$ and was fabricated using conventional dry-transfer technique\cite{Pizzocchero_2016}. All the Ohmics including the floating contact were defined using electron-beam lithography (EBL) followed by reactive-ion etching with $CHF_3$ and $O_2$. After that Cr/Pd/Au $(5/12/60nm)$ were deposited using a thermal evaporator kept at a pressure of $1-2\times10^{-7}$ mbar. The local top-gate was then fabricated with a subsequent EBL and metallization. Finally, the device geometry was patterned with EBL and etched using $CHF_3$ and $O_2$.
\\
\noindent\textbf{Measurements.} All the measurements were carried out in a cryo-free dilution refrigerator with a base temperature of $\sim20mK$. The transport measurements were performed using standard lock-in techniques.
The filling conversions from the applied gate voltages $V_{BG}$ and $V_{TG}$ at a particular magnetic field $B$, are given as follows:
\begin{align*}
\nu_{BG}=\frac{C_{BG}\phi_0}{B}(V_{BG}-V_{BG0}) \quad \text{and} \quad \nu_{TG}=\frac{C_{BG}\phi_0}{B}(V_{BG}-V_{BG0})+\frac{C_{TG}\phi_0}{B}(V_{TG}-V_{TG0})
\end{align*}
Here, $\phi_0=\frac{h}{e}$ is the magnetic flux quanta and $V_{BG0}$ and $V_{TG0}$ are the positions the charge neutrality point (CNP) of the back-gated and top-gated regions, respectively.
\\
For noise measurements, we use a resonant $LC$-tank circuit tuned around $\sim750kHz$. The signal was amplified using a homemade cryo-amplifier based on
HEMT, operating at $4K$. This is followed by further amplification at room temperature, and finally recorded using a spectrum analyzer (see detailed schematic in the SI section 1). At $I_S=0$, the amplifier measures the equilibrium voltage fluctuation given by,
\begin{equation}
S_V(0)=G^2(4k_BTR+v_n^2+i_n^2R^2)BW
\end{equation}
Here, $G$ is the total gain of the amplifier chain, $BW$ is the measurement bandwidth, and $v_n$ and $i_n$ are the voltage and current noise of the amplifier, respectively.  The first term in the equation represents the thermal noise due to the finite electronic temperature of the system. At finite bias, Joule heating in the FC elevates its temperature, adding excess thermal noise on top of $S_V(0)$. This excess noise is calculated by subtracting the zero-bias noise, $\delta S_V=S_V(I_S)-S_V(0)$. Finally, the excess current noise is obtained by, $S=\frac{\delta S_V}{R^2}$, where $R$ is the resistance seen by amplifier, given by $R=\frac{h}{\nu_{BG}e^2}$.
\\
\noindent\textbf{Theoretical model:}
We refer to \cref{Device sch.}a and assume full charge and thermal equilibration among different edges allowing us to
define local voltage and temperature at each junction of the device. Henceforth, we theoretically calculate the following quantities to
analyze the transport and excess noise in the device in the n-p-n region (see supplementary for details). Notably, the calculations are also valid for the p-n-p region
and any effect of edge reconstruction
is washed out because of full equilibration.
\newline
(1) \textit{Electrical conductance}-- We assume full charge equilibration in each
segment of the device (c.f. \cref{Device sch.}a) leading to $L \gg l^{\text{ch}}_{\text{eq}}$,
where $L$ and $l^{\text{ch}}_{\text{eq}}$ are
the geometric length and charge
equilibration length respectively.
To calculate the electrical
conductance plateau, we ground
the contact $S_2$ and bias the contact $S_1$ by a dc voltage $V_0$
corresponding to a source current $I_{S}=|\nu_{\text{BG}}| V_0 \frac{e^2}{h}$.
Electrical current conservation at each junction of the device results in
the following.
\begin{equation}   I_{D_2}=\frac{|\nu_{\text{BG}}||\nu_{\text{TG}}|}{2|\nu_{\text{BG}}|+5|\nu_{\text{TG}}|}V_0\frac{e^2}{h}, I_{D_1}=I_{\bar{D}_1}=\frac{|\nu_{\text{BG}}|(|\nu_{\text{BG}}|+2|\nu_{\text{TG}}|)}{2|\nu_{\text{BG}}|+5|\nu_{\text{TG}}|}V_0\frac{e^2}{h},
\end{equation}
where $I_{D_2}, I_{D_1}, I_{\bar{D}_1}$ are the electrical currents in drains
$D_2, D_1,\bar{D}_1$
respectively, satisfying $I_{S}=I_{D_2}+I_{D_1}+I_{\bar{D}_1}$. The potential $V_M$ of the FC is
\begin{equation}
V_\text{M}=\frac{|\nu_{\text{BG}}|+2|\nu_{\text{TG}}|}{2|\nu_{\text{BG}}|+5|\nu_{\text{TG}}|}V_0
\end{equation}
and thereby the two-terminal electrical conductance $G_\text{T}$ corresponding to the n-p-n
region is
\begin{equation}  G_\text{T}=\frac{I_{D_2}}
{V_\text{M}}=\frac{|\nu_{\text{BG}}||\nu_{\text{TG}}|}{|\nu_{\text{BG}}|+2|\nu_{\text{TG}}|}\frac{e^2}{h}.
\end{equation}
Now for $|\nu_{\text{BG}}|=2, |\nu_{\text{TG}}|=1$, we have
\begin{equation}
\frac{I_{D_2}}{I_{S}}=\frac{1}{9},\
G_\text{T}=\frac{1}{2}\frac{e^2}{h}
\end{equation}
consistent with the experimental data in \cref{Device Char.}c.
\\
(2) \textit{Thermal conductance--}
To calculate the thermal conductance, we bias both the contacts $S_1$ and $S_2$ corresponding to a source current $I_{S}$ in
$S_1$ and $-I_{S}$ in $S_2$ (c.f. \cref{Device sch.}a). This makes the effective electrochemical potential of the FC to be zero. Now, the heat balance equation dictates that the total power ($P$) dissipated at the floating contact
is equal to the total heat current ($J_Q$) flowing out of it
and therefore $P=J_Q$. Voltage drops at the FC lead to
$P=|\nu_{\text{BG}}|e^2V_0^2/h$. We denote the
number of co-propagating edge modes for $|\nu_{\text{BG}}|$
is $|n|$ and that of $|\nu_{\text{TG}}|$ is $|m|$. In the full
thermal equilibration regime,
leading to $L \gg l^{\text{th}}_{\text{eq}}$
where $l^{\text{th}}_{\text{eq}}$ is
the thermal
equilibration length,
we have
\begin{equation}
J_Q=\frac{\kappa_0}{2}\kappa_{\text{th}}(T_M^2-T_0^2),\ \kappa_{\text{th}}=\Bigg[ 2|n|+\frac{|m||n|}{2|m|+|n|} \Bigg],
\end{equation}
where $T_M$ is the temperature of the FC and $T_0$
is the temperature of all other contacts.
The term $\frac{|m||n|}{2|m|+|n|}$ arises solely
due to the n-p-n region.
Notably, for the particular case considered here, we have
$\kappa_{\text{th}}=4.5$ for $|n|=2, |m|=1$
consistent with the experimental data in
\cref{half_heat_cond}b.
\\
(3) \textit{Johnson-Nyquist noise--} To compute $T_M$ experimentally, a known prescription is to use the excess
noise. Therefore, we calculate the excess noise
$S_{D_1} (=S_{\bar{D}_1})$ and $S_{D_2}$
at drains
$D_1$ (or $\bar{D}_1$) and $D_2$, respectively, by
using the following procedure. At each junction of the device, we write the conservation equations of electrical
current fluctuation and each of which has two
contributions---a fluctuation from local voltage and a thermal fluctuation. The latter is related to the local temperature via Johnson-Nyquist noise.
Now, we self-consistently solve these equations to
calculate the excess noise at each drain.
Our calculations show that $S_{D_1} (=S_{\bar{D}_1})$ and $S_{D_2}$ depend on $T_0$ and $T_M$ in complicated ways as
\begin{equation}
\begin{split}
&S_{D_1} (=S_{\bar{D}_1})\sim f_1 \Big(T_0, T_M, \sqrt{T^2_0+T^2_M}\Big)\frac{e^2 k_B}{h},\\&
S_{D_2}\sim f_2 \Big(T_0, T_M, \sqrt{T^2_0+T^2_M}\Big)\frac{e^2 k_B}{h},
\end{split}
\end{equation}
where $f_1$ and $f_2$ depict distinct functional dependence on $T_0, T_M$ (the exact form is shown in the supplementary). Extracting $T_M$ from those
requires to solve higher-order equations. Remarkably, we find
that the a suitable combination ($S_{\text{Linear}}$) of $S_{D_1}$ and $S_{D_2}$ leads to the following expression where
$S_{\text{Linear}}$ is linearly dependent on $T_M$
and therefore
\begin{equation}
S_{\text{Linear}}=S_{D_1} - \frac{1}{4}S_{D_2}=|\nu_{\text{BG}}|\frac{e^2}{h}k_B(T_M-T_0),
\end{equation}
as shown in the experimental data in \cref{half_heat_cond}a.

\noindent\textbf{Data availability}\\
The data presented in the manuscript are available from the corresponding author upon request.

\noindent\textbf{References}
\bibliography{references}

\noindent\textbf{Acknowledgements:}
The authors thank Prof. Moty Heiblum for critical comments on the manuscript. U.R. and S.M.\ thank Arup Kumar Paul for useful discussions.
We thank the International Centre for Theoretical Sciences (ICTS) for participating in the program - Condensed Matter meets Quantum Information (code: ICTS/COMQUI2023/9), where the collaboration was initiated.
S.M.\ was supported by the Weizmann Institute of Science, Israel Deans fellowship through Feinberg Graduate School.
Y.G.\ was supported by the InfoSys Chair, IISc, Bangalore.
S.M.\ and Y. G.\ were also supported by the Minerva Foundation and grant no 2022391 from the United States--Israel Binational Science Foundation (BSF), Jerusalem, Israel.
Ankur Das was supported by IISER, Tirupati Startup grant and ANRF/ECRG/2024/001172/PMS.
M.G.\ has been supported by the Israel Science Foundation (ISF) and the Directorate for Defense Research and Development (DDR\&D) through Grant No. 3427/21, the ISF Grant No. 1113/23, and the US-Israel Binational Science Foundation (BSF) through Grant No. 2020072.
A.D. thanks the Department of Science and Technology (DST) and Science and Engineering Research Board (SERB), India, for financial support (SP/SERB-22-0387), (DST/NM/TUE/QM-5/2019), and also acknowledges funding through the Intensification of Research in High Priority Areas programme of the Science and Engineering Research Board (Grant No. IPA/2020/000034). A.D. also thanks CEFIPRA project SP/IFCP-22-0005. Growing the hBN crystals received support from the Japan Society for the Promotion of Science (KAKENHI grant nos. 19H05790, 20H00354 and 21H05233) to K.W. and T.T.
The authors gratefully acknowledge the use of \textit{Blender} (\url{https://www.blender.org/}) and \textit{Inkscape} (\url{https://inkscape.org/}) --- both free and open- source software ---- for the creation and refinement of figures presented in this work.

\noindent\textbf{Author contributions}
K.W. and T.T. synthesized the hBN crystals. U.R. contributed to device fabrication, measurement, data acquisition, and analysis. S.C. contributed to the measurement. S.M., A.D, M.G and Y.G contributed to the development of theory, analysis and
data interpretation. A.D. contributed in designing the experiment, data interpretation, and analysis.  All the authors contributed to writing the manuscript.

\noindent\textbf{Competing interests}\\
The authors declare no competing interests.



\section*{Supplementary material}

# Supplementary Information: Half-integer thermal conductance in the absence of Majorana mode

Ujjal Roy[1,*], Sourav Manna[2,*], Souvik Chakraborty[1], Kenji Watanabe[3], Takashi Taniguchi[3], Ankur Das[4], Moshe Goldstein[5], Yuval Gefen[2] and Anindya Das[1,†]

[1] Department of Physics, Indian Institute of Science, Bangalore, 560012, India.
[2] Department of Condensed Matter Physics, Weizmann Institute of Science, Rehovot 76100, Israel.
[3] National Institute of Material Science, 1-1 Namiki, Tsukuba 305-0044, Japan.
[4] Department of Physics, Indian Institute of Science Education and Research (IISER) Tirupati, Tirupati 517619, India.
[5] Raymond and Beverly Sackler School of Physics and Astronomy, Tel-Aviv University, Tel Aviv, 6997801, Israel.

# Contents

*These authors contributed equally: Ujjal Roy, Sourav Manna.
†anindya@iisc.ac.in

# 1 Experimental Setup

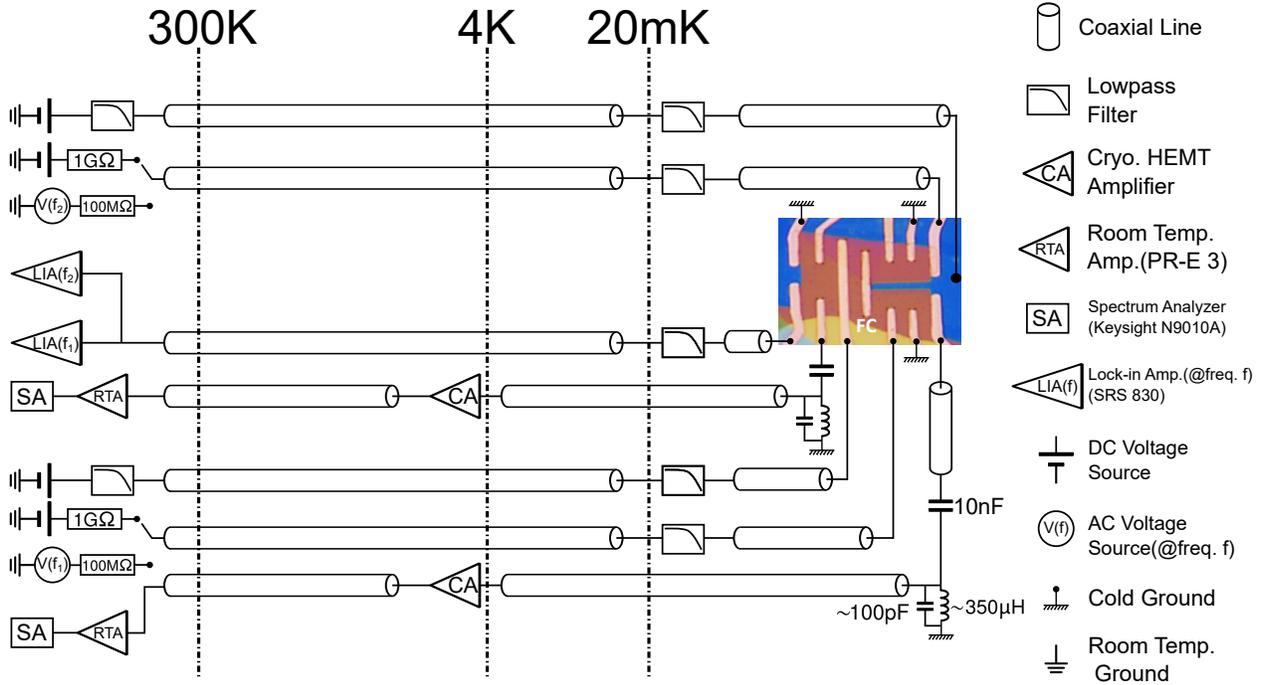

**SI Figure 1: Schematic of the detailed measurement setup.** The sample is mounted onto the mixing chamber (MC) plate of a dilution refrigerator. The gate control, current injection and voltage measurement lines are filtered by RC filters fitted at the MC plate. The cold ground(CG) lines are anchored to the cold finger. The resonant LC network is formed by a superconducting coil of $\sim 350 \mu H$ in combination with a parasitic capacitance of $\sim 100 pF$ contributed by the coaxial line running from the MC plate to the input of the cryo-amplifier(CA) situated at the $4K$ plate.

## 2 Gain Calibration and $T_0$ determination

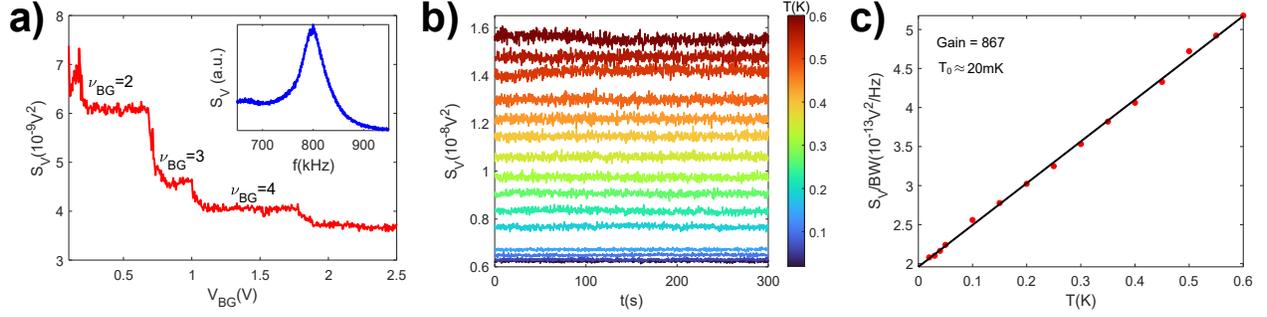

**SI Figure 2: Gain calibration and $T_0$ determination at contact $D_2$** **(a)** $S_V(0)$ (as mentioned in Eq.3 of the main text) as a function of backgate voltage measured at $f_0 = 798kHz$, which is shown in the inset. **(b)** Traces of $S_V(0)$ with time measured at the center of $\nu_{BG} = 2$ plateau and at $f_0$ for bath temperature varying from $T = 20mK$ to $600mK$. The colorbar represents the temperature of the corresponding trace. **(c)** The time average of $S_V$ from **(b)** divided by the measurement bandwidth ($BW = 30kHz$) plotted against the bath temperature $T$ (red solid circles). As from Eq.3 of the main text and also from our earlier works[1-4] we can extract gain using $G = \sqrt{\frac{\partial S_V/\partial T}{4k_B R}}$, this yields a gain value of 867. The data measured at the base temperature ($T = 20, mK$) aligns well with the linear fit (solid black line), indicating the value of the base electronic temperature $T_0$.

# 3 Additional data for Device 1: Characterization

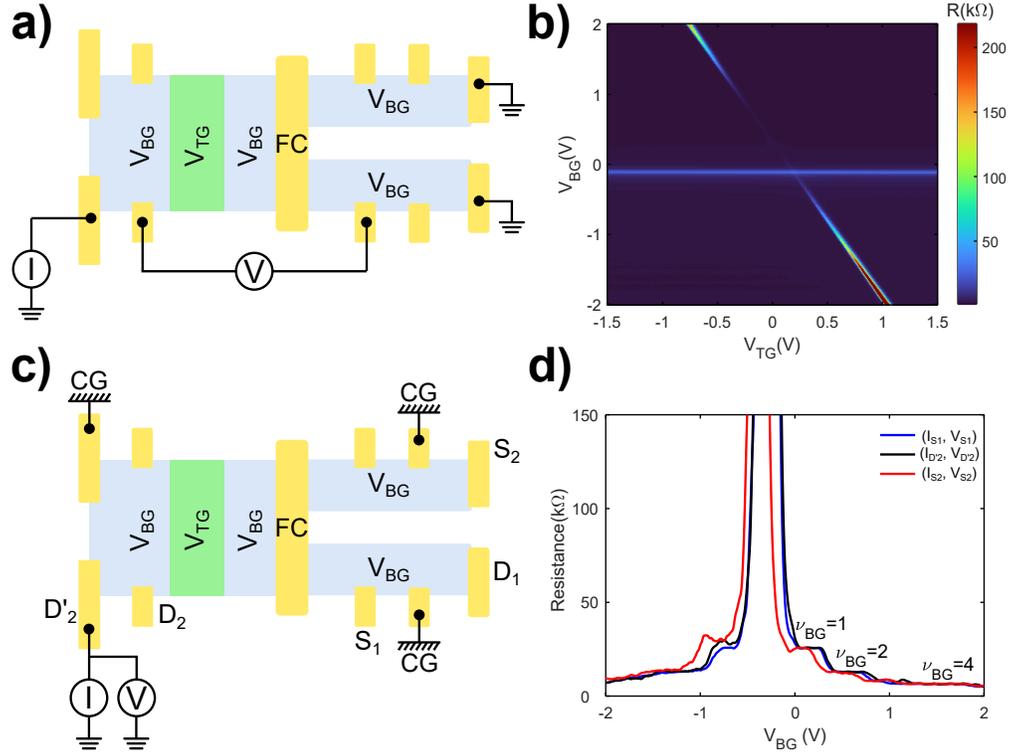

**SI Figure 3: Device 1 characterization.** (a) Schematic of the measurement setup used to characterize the device at magnetic field, $B = 0$. (b) 2D colormap of the 4-probe resistance in $V_{TG} - V_{BG}$ plane at $B = 0$ and $T = 300 mK$, following the schematic as shown in (a), using standard lock-in techniques with an ac excitation of $5nA$. The colorbar represents the resistance values. The high resistance strip parallel to $V_{TG}$ axis marks the charge neutrality point (CNP) of the bulk BLG region, while the diagonal insulating band corresponds the CNP of the top gated region. The increasing resistance with electric field indicates the expected band gap opening in BLG. (c) Setup used to characterize the quantum Hall (QH) response of the individual arms, with at least one cold ground present in each arm. (d) 2-probe QH response in each of the arm as function $V_{BG}$ taken at $B = 6T$ and $T = 20mK$, measured using the configuration shown in (c). Well-developed plateaus at $\nu = 1, 2$ and $4$ are observed across all arms, with good overlap, meeting a key requirement for the experiment.

## 4 Additional data for Device 1: Transmission from different source contacts and simplified conductor model

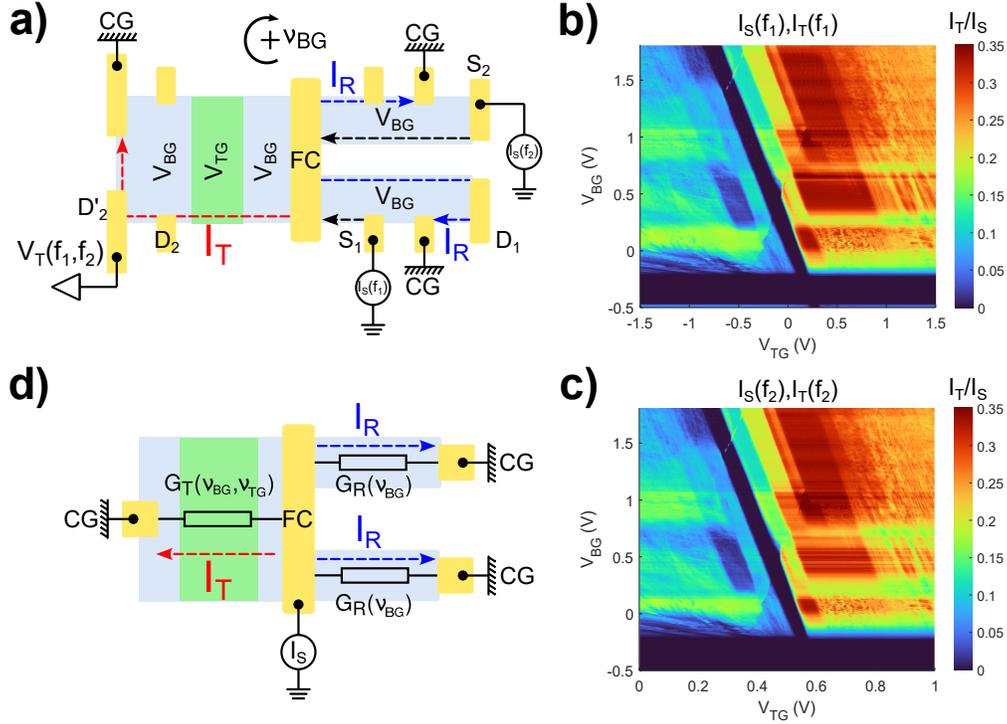

**SI Figure 4:** $S_1$ **&** $S_2$ **Response and simplified conductor model.** **(a)** Setup used to measure the transmitted current $I_T$, also shown in fig. 2(a) of the main manuscript. Here 5 nA of current of different frequencies, marked as $f_1$ and $f_2$ are injected at $S_1$ and $S_2$ simultaneously. As discussed in the main manuscript the injected current will divide from FC. The voltage drop $V_T$ resulting from the transmitted current is detected at the corresponding frequencies. The $I_T(S_i)$, resulting as current injected from source $S_i$ is given by $I_T(S_i) = \frac{V_T(f_i)}{R_{D'_2}}$. Here $R_{D'_2}$ is the resistance measured at $D'_2$ at corresponding $V_{BG}$, as shown in Supplementary Fig. 1(d). **(b)-(c)** 2D colormap of the $I_T/I_S$ resulting from current injected at $S_1$ and $S_2$ respectively, as function of $V_{BG}$ and $V_{TG}$. These responses are identical, a crucial requirement for this experiment. From here we also extract $V_{BG0} = -0.31V$ and $V_{TG0} = 0.21V$, the CNPs of the topgated and backgated region, which are used to evaluate the respective filling factors (see the 'Methods' section of the main manuscript). Same plot is shown in Fig. 2(b) of the main manuscript plotted in terms of respective filling factors. **(d)** The response of our three arm device can be very easily understood with the simplified conductor model. Three conductors replaced by the respective arms are connected parallelly between the FC and CG. The conductance values are discussed in the main manuscript. The injected current divides according to the conductance of the individual arms.

# 5 Additional data for Device 1: Transmission current at different magnetic fields and equilibration mechanism

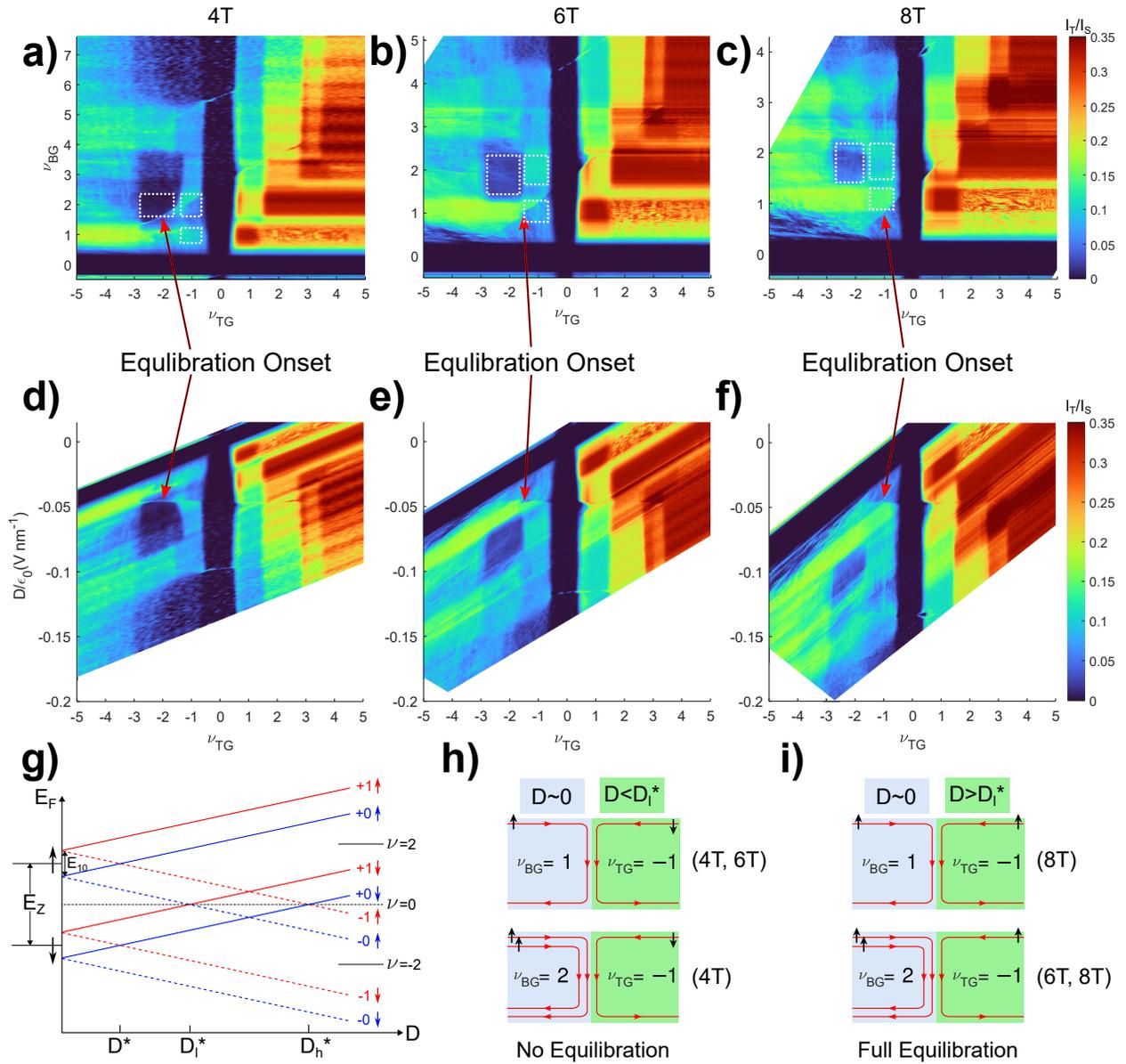

**SI Figure 5: Degree of equilibration at different B. (a)-(c)** 2D colormap of measured $I_T/I_S$ as a function of $\nu_{BG}$ and and $\nu_{BG}$ for $B = 4T$ **(a)**, $B = 6T$ **(b)** and $B = 8T$ **(c)**. A distinct transition in the current ratio — marked by a red arrow — indicates the onset of full equilibration. White dashed rectangles highlight key filling factor configurations: $(\nu_{BG}, \nu_{TG}) = (1, -1)$, $(2, -1)$ and $(2, -2)$. At $B = 4T$, all highlighted regions exhibit only partial equilibration. At $B = 6T$, full equilibration is achieved in the $(2, -1)$ region, while at $B = 8T$, both $(1, -1)$ and $(2, -1)$ regions show full equilibration. However, the $(2, -2)$ region consistently shows suppressed $I_T$, indicating the absence of full equilibration. **(d)-(f)** Corresponding plots of $I_T/I_S$ versus $\nu_{TG}$ and $D$, where $D = \frac{C_{TG}(V_{TG}-V_{TG0})-C_{BG}(V_{BG}-V_{BG0})}{2}$, the displacement field of the topgated region. A clear transition point can be seen with electric field ($D/\epsilon_0 \approx -0.05 V/nm$) beyond which full equilibration can be observed, which we marked as 'Equilibration Onset'( pointed by the red arrows). **(g)** An effective diagram of ZLL of BLG at a fixed magnetic field with out of plane displacement field, adapted from[5–7]. Here $E_Z$ is the Zeeman energy and $E_{10}$ is the energy spacing between the zeroth and first orbital of BLG at that magnetic field. Transition fields $D^*$, $D_l^*$ and $D_h^*$ are also marked following the conventions of Refs.[5,6]. **(h)** Scenario for no equilibration for $(\nu_{BG}, \nu_{TG}) = (1, -1)$ (top panel) and $(2, -1)$ (bottom panel). While the backgated region (light blue) always remains at $D \sim 0$, the electric field of the topgate control region (light green) can vary. For $D < D_l^*$, the spin configurations of the copropagating edges remain orthogonal according to the diagram **(g)**, preventing equilibration. However, we observe partial equilibration in our system. **(i)** Spin configurations when $D > D_l^*$, where spin states align, leading to full equilibration — consistent with the experimental data. However persistent suppression of $I_T$ for $(2, -2)$ requires beyond spin-selective equilibration model.

# 6 Additional data for Device 1: Thermal noise in unipolar region, taken at $D_2$

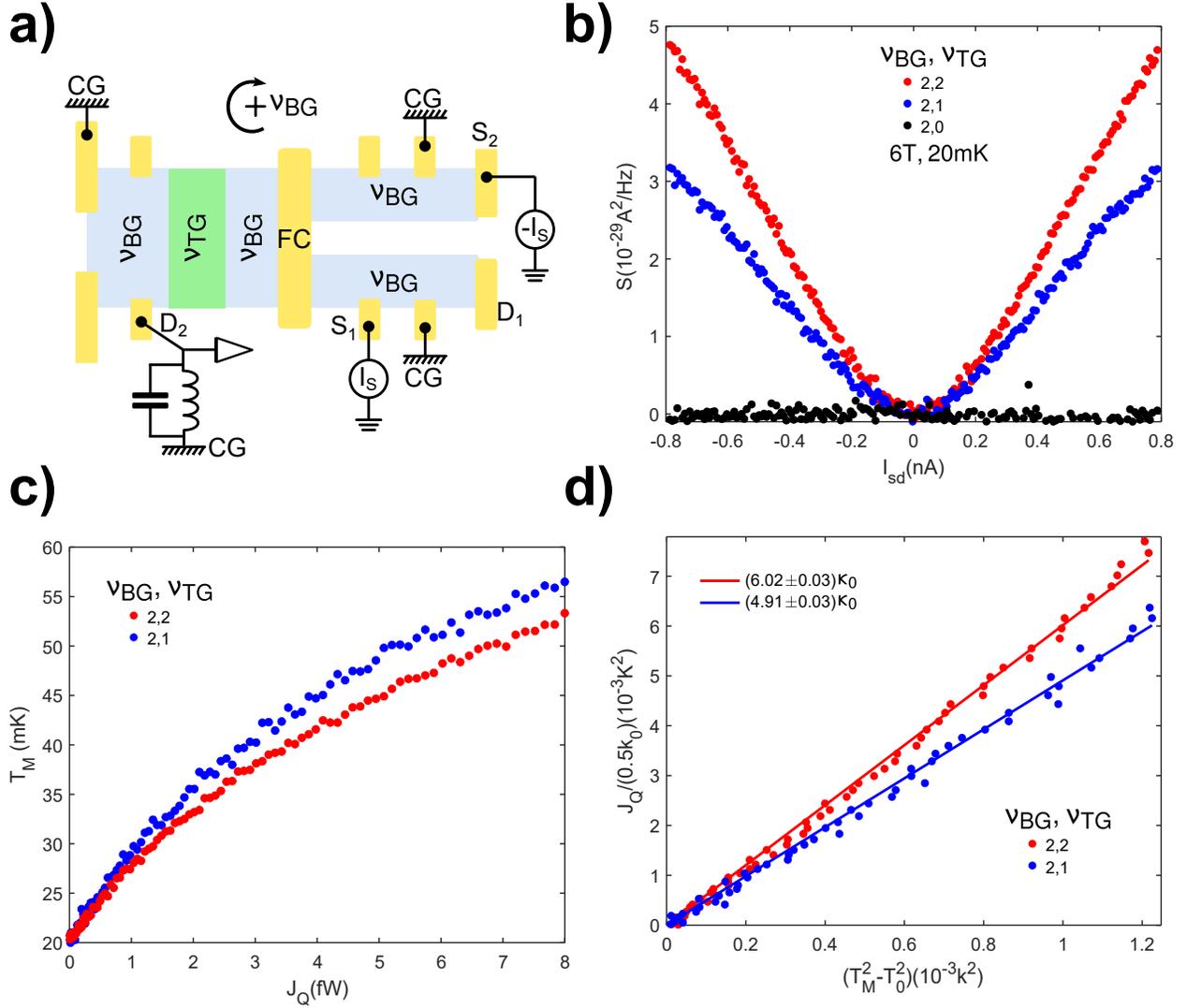

**SI Figure 6: Thermal noise in unipolar region, taken at $D_2$.** (a) Schematic used for noise measurement at detector $D_2$, details mentioned in Fig.1(a) of the main manuscript. (b) Excess thermal noise measured at $D_2$ as a function of injected DC current $I_S$ for three configurations: $(\nu_{BG}, \nu_{TG}) = (2, 2)$ (red circles), $(2, 1)$ (blue circles), and $(2, 0)$ (black circles), at $B = 6T$ and $T = 20mK$. For $(2, 0)$, the top-gated region becomes insulating, resulting in no observable excess noise. (c) Extracted electronic temperatures $T_M$ of the floating contact, obtained from panel (b), plotted as a function of the dissipated power $J_Q = P$ due to Joule heating. (d) Normalized power dissipation $J_Q/(0.5\kappa_0)$ plotted against $(T_M^2 - T_0^2)$ for configurations $(2, 2)$ and $(2, 1)$. The solid lines are linear fits used to extract the thermal conductance. The extracted values are $\kappa_{th} = 6.02\kappa_0$ for $(2, 2)$ and $4.91\kappa_0$ for $(2, 1)$, as expected.

# 7 Additional data for Device 1: 1/2-integer thermal conductance at $7T$ and $8T$

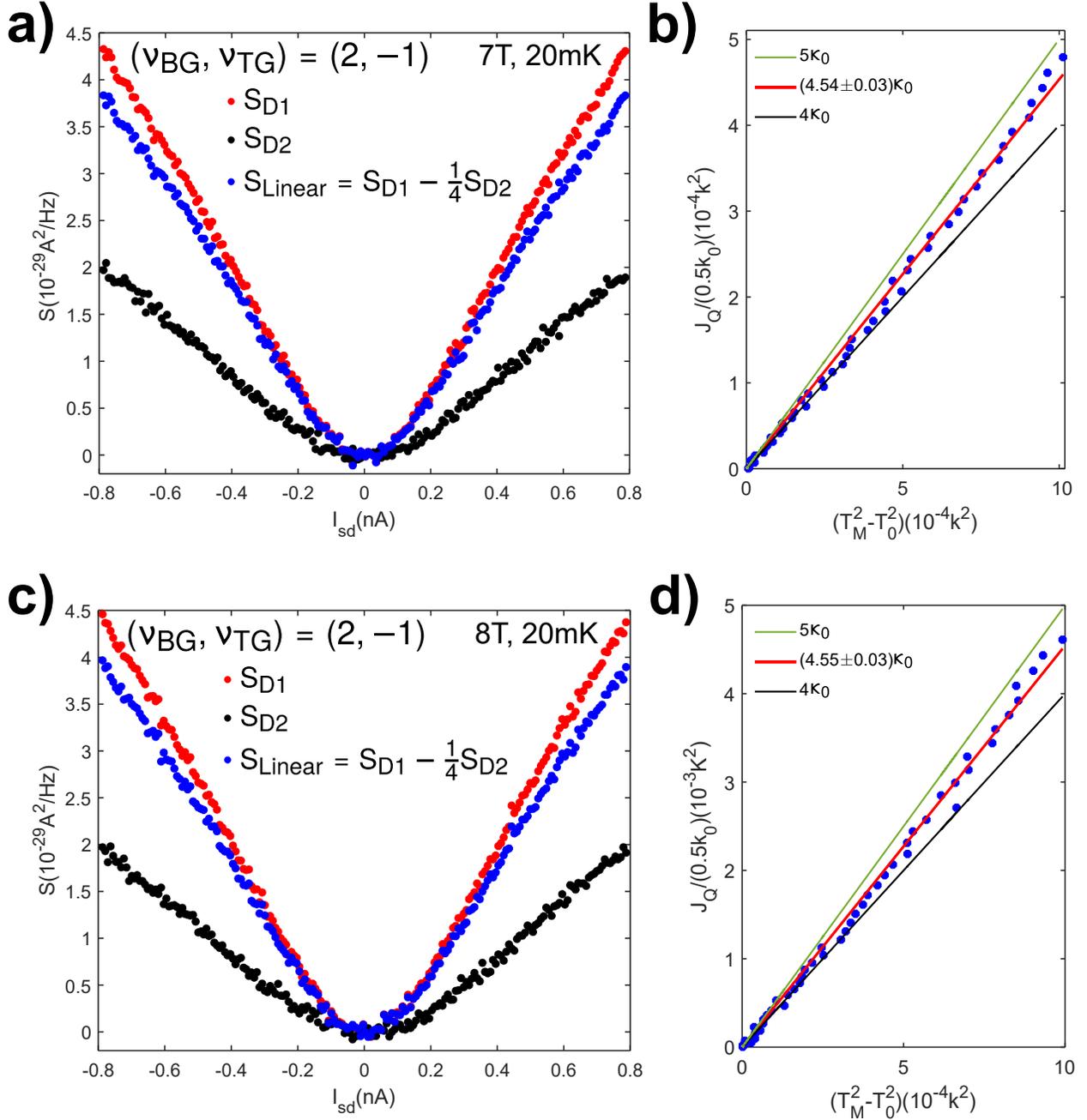

**SI Figure 7:** 1/2-**integer thermal conductance at** $7T$ **and** $8T$. **(a),(c)** Excess thermal noise measured at the half-conductance plateau $(\nu_{BG}, \nu_{TG}) = (2, -1)$, at $D_1$(red circles, $S_{D1}$) and at $D_2$(black Circles, $S_{D2}$) as a function of $I_S$. $S_{Linear} = S_{D1} - \frac{1}{4}S_{D2}$ is shown by the blue circles, for $B = 7T$ **(a)** and $B = 8T$ **c**. **(b),(d)** $\Delta J_Q/0.5\kappa_0$ is plotted(solid circles) as a function of $T_M^2 - T_0^2$. The slope of the linear fitting(solid line) gives the thermal conductance value of $4.54\kappa_0$ **(b)** and $4.62\kappa_0$ **(d)**.

# 8 Data for Device 2: Characterization

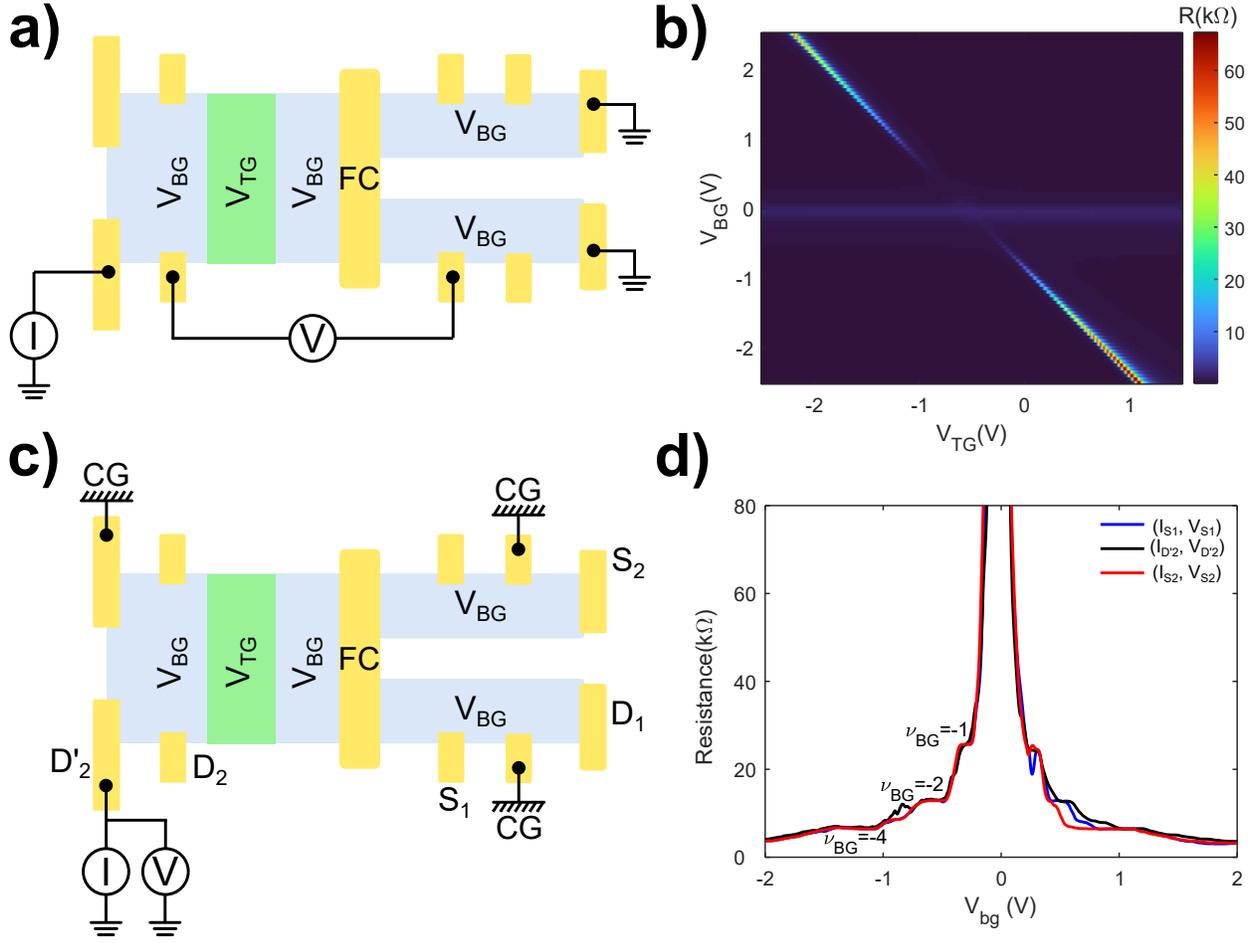

**SI Figure 8: Device 2 characterization** (a) Schematic of the measurement setup used to characterize the device at magnetic field, $B = 0$. (b) 2D colormap of the 4-probe resistance in $V_{TG} - V_{BG}$ plane at $B = 0$ and $T = 4K$, following the schematic as shown in (a), using standard lock-in techniques with an ac excitation of $5nA$. The colorbar represents the resistance values. The high resistance strip parallel to $V_{TG}$ axis marks the charge neutrality point(CNP) of the bulk BLG region, while the diagonal insulating band corresponds the CNP of the top gated region. The increasing resistance with electric field indicates the expected band gap opening in BLG. (c) Setup used to characterize the quantum Hall(QH) response of the individual arms, with at least one cold ground present in each arm. (d) 2-probe QH response in each of the arm as function $V_{BG}$ taken at $B = 6T$ and $T = 20mK$, measured using the configuration shown in (c). Well-developed plateaus at $\nu = -1, -2$ and $-4$ are observed across all arms, with good overlap, meeting a key requirement for the experiment.

# 9 Data for Device 2: 1/2-integer electrical conductance

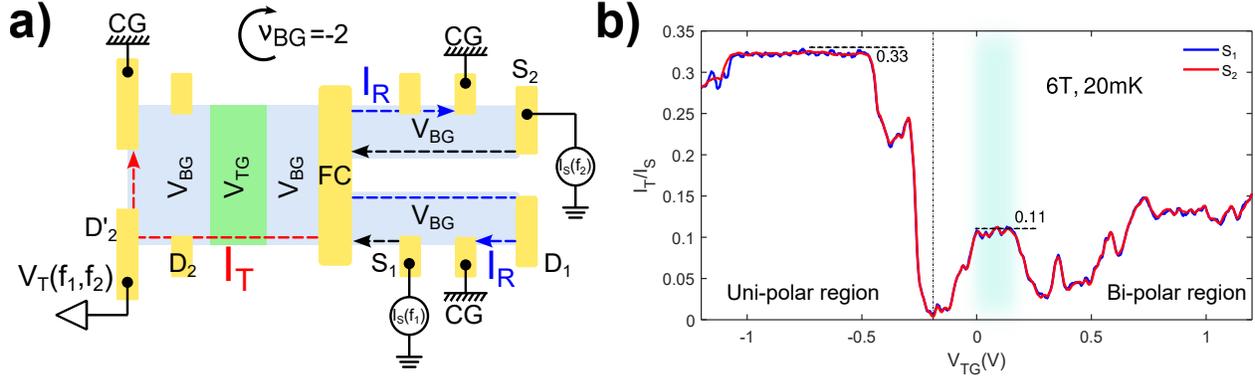

**SI Figure 9:** $S_1$ & $S_2$ **Response (a)** Schematic for measuring $I_T$, details are mentioned in SI Fig.2(a). **(b)** $I_T/I_S$ as a function $V_{TG}$ at $\nu_{BG} = 2$ ($V_{BG} = -0.57$), while the current injected at $S_1$(blue trace) and $S_2$(red trace). Response from the two sources are identical. For $(\nu_{BG}, \nu_{TG}) = (-2, -2)$, $I_T/I_s \approx 0.33$ satisfying equipartition of current at FC and for $(-2, 1)$, $I_T/I_s \approx 0.11$ indicating full equilibration.

## 10 Data for Device 2: Thermal conductance in unipolar region

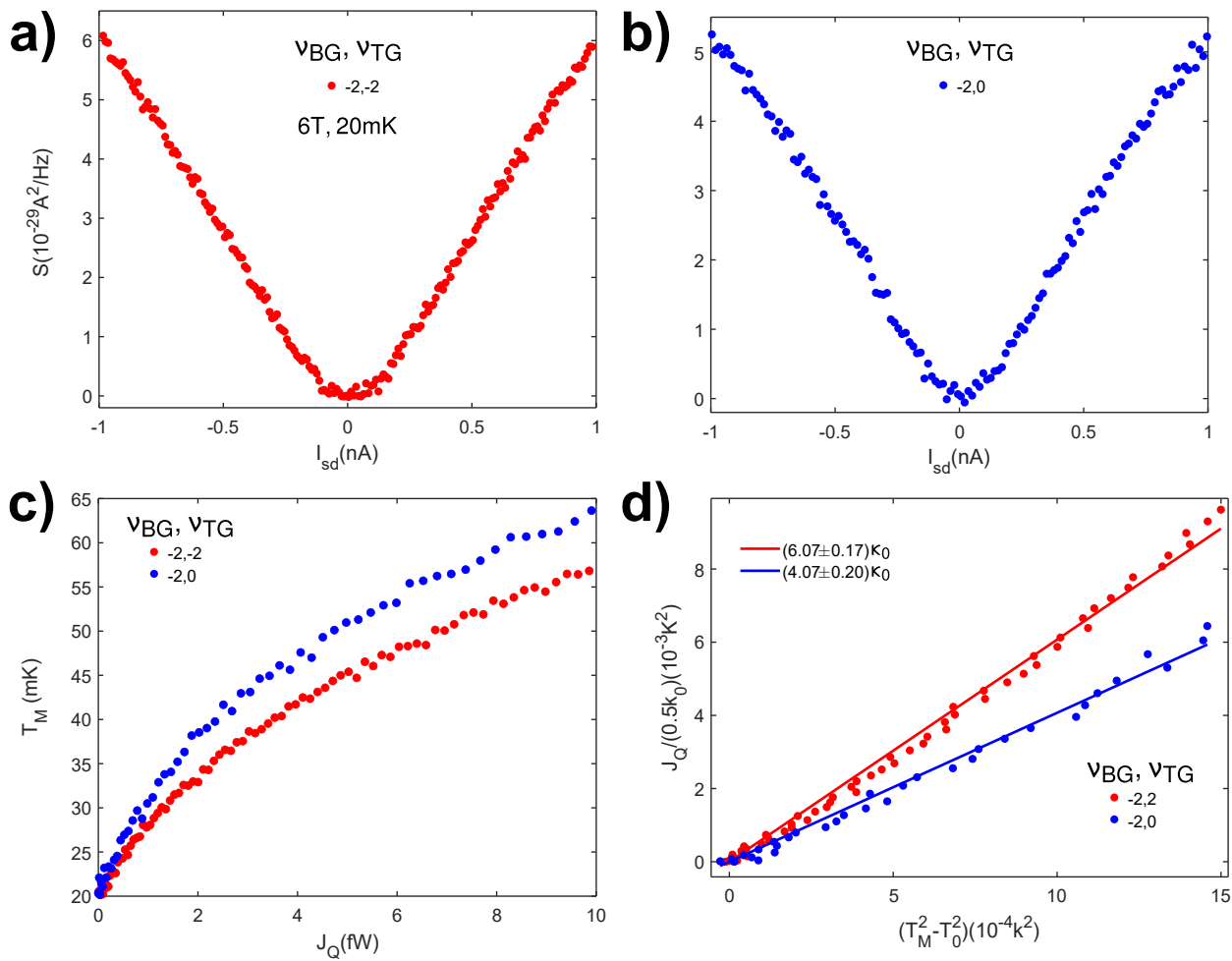

**SI Figure 10:** Thermal noise in unipolar region, taken at $D_1$.

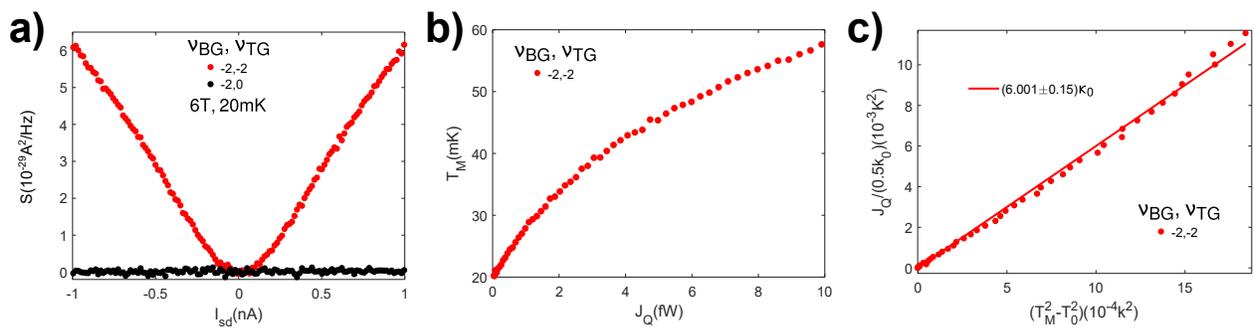

**SI Figure 11:** Thermal noise in unipolar region, taken at $D_2$.

# 11  Data for Device 2: $1/2$-integer thermal conductance

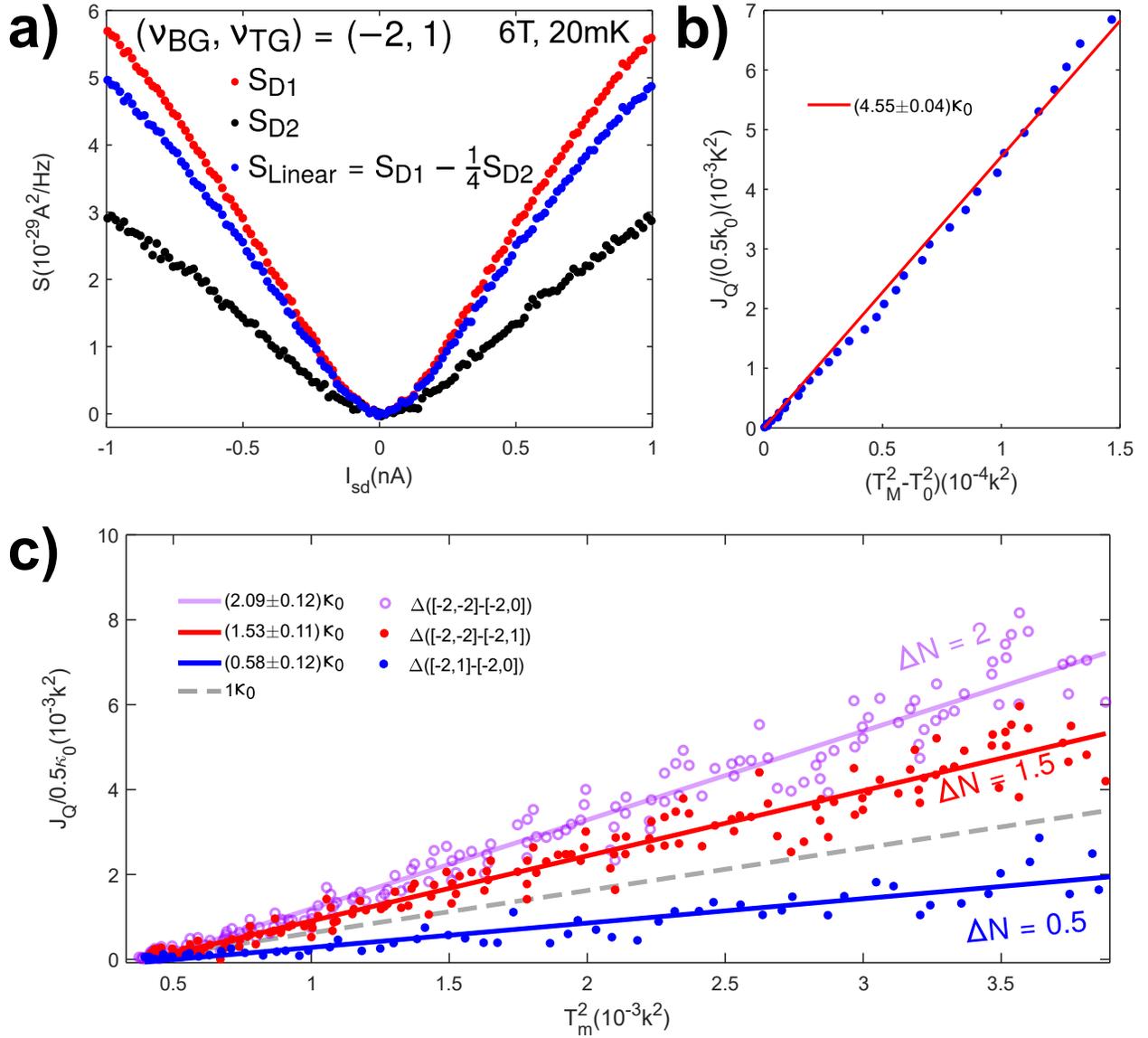

**SI Figure 12: $1/2$-integer thermal conductance.** (a) Excess thermal noise measured at the half-conductance plateau $(\nu_{BG}, \nu_{TG}) = (-2, 1)$, at $D_1$(red circles, $S_{D1}$) and at $D_2$(black Circles, $S_{D2}$) as a function of $I_S$. $S_{Linear} = S_{D1} - \frac{1}{4}S_{D2}$ is shown by the blue circles. (b) $\Delta J_Q/0.5\kappa_0$ is plotted(solid circles) as a function of $T_M^2 - T_0^2$. The slope of the linear fitting(solid line) gives the thermal conductance value of $4.55\kappa_0$. (c) $\Delta J_Q/0.5\kappa_0$ is plotted against $T_M^2$ for $\Delta N = 2$ (between $[-2, -2]$ & $[-2, 0]$), $\Delta N = 1.5$ (between $[-2, -2]$ & $[-2, 1]$), and $\Delta N = 0.5$ (between $[-2, 1]$ & $[-2, 0]$), where $\Delta J_Q = J_Q(N_i, T_M) - J_Q(N_j, T_M)$. The solid lines are linear fittings to extract the the thermal conductance associated with $\Delta N$ 1D channels. The extracted values are $2.09\kappa_0$, $1.53\kappa_0$ and $0.58\kappa_0$ for $\Delta N = 2$, $\Delta N = 1.5$ and $\Delta N = 0.5$ respectively. The dashed grey line represents the theoretical $1\kappa_0$ line expected for $\Delta N = 1$.

13

## 12 Derivations of the theoretical equations

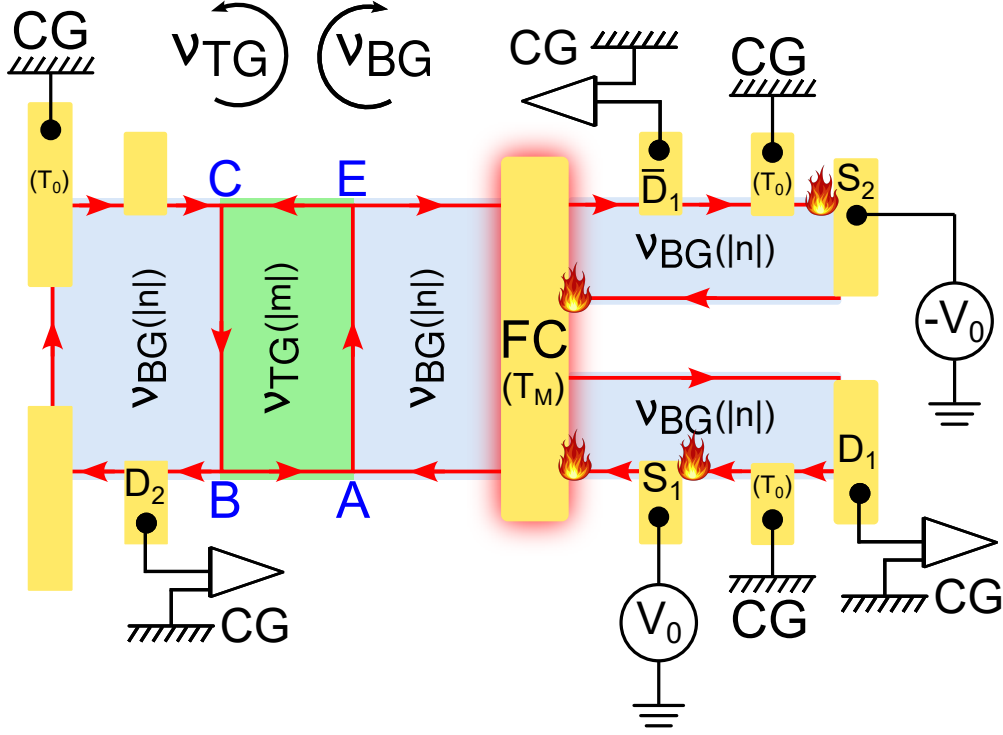

**SI Figure 13: Schematic of the device.** Two sources $S_1, S_2$ are oppositely biased and $D_1, \bar{D}_1, D_2$ are the drains. Powers are dissipated in the hotspots (depicted by fire signs) at the floating contact (FC) raising its temperature to $T_M$, while the cold grounds (CG) are at temperature $T_0$. Bottom gate filling ($\nu_{\mathrm{BG}}$ with $n$ number of co-propagating modes) and top gate filling ($\nu_{\mathrm{TG}}$ with $m$ number of co-propagating modes) have opposite chiralities, as shown by the circular arrows.

We consider full charge and thermal equilibration at each segment of the device (Fig. 13). Thereby, $L \gg l_{\mathrm{eq}}^{\mathrm{ch}}$ and $L \gg l_{\mathrm{eq}}^{\mathrm{th}}$, where $L$ is the geometric length of the segment and $l_{\mathrm{eq}}^{\mathrm{ch}}$ and $l_{\mathrm{eq}}^{\mathrm{th}}$ are charge and thermal equilibration lengths respectively. Such equilibration process between two chiral modes is modeled by employing a number of virtual reservoirs that neither take any charge nor any energy from the system. In a steady-state, we neglect any temporary accumulation of charge or energy in the reservoirs. We define local voltage and temperature at each junction of our set up (Fig. 13) and explicitly calculate the transport properties and excess noise as follows.

**Electrical conductance.** We refer to Fig. 13 and consider a ground at contact $S_2$. We assume full charge equilibration at each segment and write the following electrical current conservation equations at $A, E, C, B$,

14

and at the floating contact ($M$), respectively.

$$\begin{aligned}
|\nu_{\text{TG}}|V_B + |\nu_{\text{BG}}|V_M &= V_A(|\nu_{\text{BG}}| + |\nu_{\text{TG}}|), \\
V_A(|\nu_{\text{BG}}| + |\nu_{\text{TG}}|) &= V_E(|\nu_{\text{BG}}| + |\nu_{\text{TG}}|), \\
|\nu_{\text{TG}}|V_E + 0 &= V_C(|\nu_{\text{BG}}| + |\nu_{\text{TG}}|), \\
V_C(|\nu_{\text{BG}}| + |\nu_{\text{TG}}|) &= V_B(|\nu_{\text{BG}}| + |\nu_{\text{TG}}|), \\
|\nu_{\text{BG}}|V_0 + 0 + |\nu_{\text{BG}}|V_E &= 3|\nu_{\text{BG}}|V_M.
\end{aligned} \tag{S1}$$

Solving those equations self-consistently, we find

$$\begin{aligned}
V_M &\to \frac{V_0(|\nu_{\text{BG}}| + 2|\nu_{\text{TG}}|)}{2|\nu_{\text{BG}}| + 5|\nu_{\text{TG}}|}, V_A \to \frac{V_0(|\nu_{\text{BG}}| + |\nu_{\text{TG}}|)}{2|\nu_{\text{BG}}| + 5|\nu_{\text{TG}}|}, V_E \to \frac{V_0(|\nu_{\text{BG}}| + |\nu_{\text{TG}}|)}{2|\nu_{\text{BG}}| + 5|\nu_{\text{TG}}|}, \\
V_C &\to \frac{|\nu_{\text{TG}}|V_0}{2|\nu_{\text{BG}}| + 5|\nu_{\text{TG}}|}, V_B \to \frac{|\nu_{\text{TG}}|V_0}{2|\nu_{\text{BG}}| + 5|\nu_{\text{TG}}|}.
\end{aligned} \tag{S2}$$

The electrical current at each drain can be calculated as

$$I_{D_2} = |\nu_{\text{BG}}|V_B \frac{e^2}{h}, I_{\bar{D}_1} = |\nu_{\text{BG}}|V_M \frac{e^2}{h}, I_{D_1} = |\nu_{\text{BG}}|V_M \frac{e^2}{h} \tag{S3}$$

leading to

$$I_{D_2} = \frac{|\nu_{\text{BG}}||\nu_{\text{TG}}|}{2|\nu_{\text{BG}}| + 5|\nu_{\text{TG}}|} V_0 \frac{e^2}{h}, I_{D_1} = I_{\bar{D}_1} = \frac{|\nu_{\text{BG}}|(|\nu_{\text{BG}}| + 2|\nu_{\text{TG}}|)}{2|\nu_{\text{BG}}| + 5|\nu_{\text{TG}}|} V_0 \frac{e^2}{h}. \tag{S4}$$

**Thermal conductance.** We refer to Fig. 13 and bias both the contacts $S_1$ and $S_2$ corresponding to a source current $I_S$ in $S_1$ and $-I_S$ in $S_2$ making the effective electrochemical potential of the floating contact to be zero. We assume full thermal equilibration at each segment and write the following heat current conservation equations at $A, C, B, E$, respectively.

$$\begin{aligned}
\frac{1}{2}\kappa_0(|m|T_B^2) + \frac{1}{2}\kappa_0(|n|T_M^2) &= \frac{1}{2}\kappa_0(T_A^2(|m| + |n|)), \\
\frac{1}{2}\kappa_0(|m|T_E^2) + \frac{1}{2}\kappa_0(|n|T_0^2) &= \frac{1}{2}\kappa_0(T_C^2(|m| + |n|)), \\
\frac{1}{2}\kappa_0(T_C^2(|m| + |n|)) &= \frac{1}{2}\kappa_0(T_B^2(|m| + |n|)), \\
\frac{1}{2}\kappa_0(T_A^2(|m| + |n|)) &= \frac{1}{2}\kappa_0(T_E^2(|m| + n)).
\end{aligned} \tag{S5}$$

Solving those equations self-consistently, we find

$$\begin{aligned}
T_A^2 &\to \frac{|m|(T_0^2 + T_M^2) + |n|T_M^2}{2|m| + |n|}, T_B^2 \to \frac{|m|(T_0^2 + T_M^2) + |n|T_0^2}{2|m| + |n|}, \\
T_C^2 &\to \frac{|m|(T_0^2 + T_M^2) + |n|T_0^2}{2|m| + |n|}, T_E^2 \to \frac{|m|(T_0^2 + T_M^2) + |n|T_M^2}{2|m| + |n|}.
\end{aligned} \tag{S6}$$

The total heat current, flowing out of the floating contact, can be calculated as

$$J_{\text{total}} = \frac{\kappa_0}{2}|n|(3T_M^2 - 2T_0^2 - T_E^2) \tag{S7}$$

15

leading to

$$J_{\text{total}} = \frac{\kappa_0}{2}\left[2|n| + \frac{|m||n|}{2|m|+|n|}\right](T_M^2 - T_0^2). \tag{S8}$$

**Excess noise.** We refer to Fig. 13 and write the following equations depicting electrical current fluctuations at each segment.

$$\begin{aligned}
&\delta I_{BA} + \delta I_{MA} = \delta I_{AE},\ \delta I_{CB} = \delta I_{BA} + \delta I_{BD_2},\ \delta I_{EC} + \delta I_{G_3C}^{th} = \delta I_{CB}, \\
&\delta I_{AE} = \delta I_{EC} + \delta I_{EM},\ \delta I_{EM} + \delta I_{S_1M}^{th} + \delta I_{S_2M}^{th} = \delta I_{MA} + \delta I_{M\bar{D}_1} + \delta I_{MD_1}, \\
&\delta I_{BA} = \delta I_{BA}^{th} + \delta V_B \frac{e^2}{h}|\nu_{\text{TG}}|,\ \delta I_{BD_2} = \delta I_{BD_2}^{th} + \delta V_B \frac{e^2}{h}|\nu_{\text{BG}}|, \\
&\delta I_{EC} = \delta I_{EC}^{th} + \delta V_E \frac{e^2}{h}|\nu_{\text{TG}}|,\ \delta I_{EM} = \delta I_{EM}^{th} + \delta V_E \frac{e^2}{h}|\nu_{\text{BG}}|, \\
&\delta I_{M\bar{D}_1} = \delta I_{M\bar{D}_1}^{th} + \delta V_M \frac{e^2}{h}|\nu_{\text{BG}}|,\ \delta I_{MA} = \delta I_{MA}^{th} + \delta V_M \frac{e^2}{h}|\nu_{\text{BG}}|,\ \delta I_{MD_1} = \delta I_{MD_1}^{th} + \delta V_M \frac{e^2}{h}|\nu_{\text{BG}}|,
\end{aligned} \tag{S9}$$

where $\delta I_{BA}$ is the fluctuation in the electrical current flowing along the segment $BA$, and similarly others. The voltage fluctuations at $B, E, M$ are $\delta V_B, \delta V_E, \delta V_M$, respectively. We have the thermal fluctuations following the Johnson-Nyquist noise relations as

$$\begin{aligned}
&\delta^2 I_{G_3C}^{th} = 2\frac{e^2}{h}k_B|\nu_{\text{BG}}|T_0,\ \delta^2 I_{S_1M}^{th} = 2\frac{e^2}{h}k_B|\nu_{\text{BG}}|T_0, \\
&\delta^2 I_{S_2M}^{th} = 2\frac{e^2}{h}k_B|\nu_{\text{BG}}|T_0,\ \delta^2 I_{BA}^{th} = 2\frac{e^2}{h}k_B|\nu_{\text{TG}}|T_B, \\
&\delta^2 I_{BD_2}^{th} = 2\frac{e^2}{h}k_B|\nu_{\text{BG}}|T_B,\ \delta^2 I_{EC}^{th} = 2\frac{e^2}{h}k_B|\nu_{\text{TG}}|T_E, \\
&\delta^2 I_{EM}^{th} = 2\frac{e^2}{h}k_B|\nu_{\text{BG}}|T_E,\ \delta^2 I_{M\bar{D}_1}^{th} = 2\frac{e^2}{h}k_B|\nu_{\text{BG}}|T_M, \\
&\delta^2 I_{MA}^{th} = 2\frac{e^2}{h}k_B|\nu_{\text{BG}}|T_M,\ \delta^2 I_{MD_1}^{th} = 2\frac{e^2}{h}k_B|\nu_{\text{BG}}|T_M.
\end{aligned} \tag{S10}$$

Solving the equations self-consistently, we find the excess auto-correlation noise (that is subtracting the background noise $2\frac{e^2}{h}k_B|\nu_{\text{BG}}|T_0$) at the drains as

$$S_{\bar{D}_1} \equiv \delta^2 I_{M\bar{D}_1} = 2\frac{e^2}{h}k_B|\nu_{\text{BG}}| \times$$
$$\frac{\left(-2T_0(|\nu_{\text{BG}}| + 2|\nu_{\text{TG}}|)(|\nu_{\text{BG}}| + 4|\nu_{\text{TG}}|) + |\nu_{\text{TG}}|(|\nu_{\text{BG}}| + |\nu_{\text{TG}}|)(T_B + T_E) + 2T_M\left(|\nu_{\text{BG}}|^2 + 5|\nu_{\text{BG}}||\nu_{\text{TG}}| + 7|\nu_{\text{TG}}|^2\right)\right)}{(2|\nu_{\text{BG}}| + 5|\nu_{\text{TG}}|)^2},$$

$$S_{D_2} \equiv \delta^2 I_{BD_2} = 4\frac{e^2}{h}k_B|\nu_{\text{BG}}||\nu_{\text{TG}}| \times$$
$$\frac{\left(-T_0(4|\nu_{\text{BG}}| + 7|\nu_{\text{TG}}|) + 2T_B(|\nu_{\text{BG}}| + |\nu_{\text{TG}}|) + 2|\nu_{\text{BG}}|T_E + 2|\nu_{\text{TG}}|T_E + 3|\nu_{\text{TG}}|T_M\right)}{(2|\nu_{\text{BG}}| + 5|\nu_{\text{TG}}|)^2},$$

$$S_{D_1} = S_{\bar{D}_1}.$$
(S11)

Remarkably, we have found the following relation to be linearly dependent on $T_M$ as

$$S_{\text{Linear}} = S_{D_1} - \frac{1}{4}S_{D_2} = |\nu_{\text{BG}}|\frac{e^2}{h}k_B(T_M - T_0). \tag{S12}$$



\end{document}